\documentclass[useAMS,usenatbib]{mn2e}
\usepackage{graphicx,color,epsfig,amsmath,amssymb}

\title[GSPH with viscosity]{ GodunovSPH with shear viscosity : implementation and tests}
\author[Cha \& Wood]{Seung-Hoon Cha\thanks{E-mail:
seung-hoon.cha@tamuc.edu} and Matt A. Wood\\
Department of Physics \& Astronomy, Texas A\&M University-Commerce, 2600 South Neal Street, Commerce, TX 75428 USA}
\begin{document}

\date{Accepted ????. Received ????; in original form 2015 September 29 }

\pagerange{\pageref{firstpage}--\pageref{lastpage}} \pubyear{2015}

\maketitle

\label{firstpage}

\begin{abstract}
The acceleration and energy dissipation terms due to the shear viscosity
have been implemented
and tested in GodunovSPH. The double summation method has been employed
to avoid the well known numerical noise of the second derivative in particle based codes.
The plane Couette flow with various initial and boundary conditions have been
used as tests, and the numerical and analytical results show a
good agreement. Not only the viscosity--only calculation, but the full hydrodynamics simulations
have been performed, and they show expected results as well.
The very low kinematic viscosity simulations show
a turbulent pattern when the Reynolds number exceeds $\sim$$10^2$. The critical value
of the Reynolds number at the transition point of the laminar and turbulent flows coincides with the previous
works approximately. A smoothed dynamic viscosity has been suggested to describe
the individual kinematic viscosity of particles.
The infinitely extended Couette flow which has two layers of different viscosities
has been simulated to check the smoothed dynamic viscosity,
and the result agrees well with the analytic solution.
In order to compare the standard SPH and GodunovSPH, the two layers test has been performed again
with a density contrast. GodunovSPH
shows less dispersion than the standard SPH, but there is no significant difference in the results.
The results of the viscous ring evolution has also been presented as well, and the
numerical results agrees with the analytic solution.
\end{abstract}


\section{Introduction}
\label{sec:intro}
Smoothed Particle Hydrodynamics (hereafter SPH) is a widely used
method in numerical astronomy since its first announcement 
\citep{Gingold1977a,Lucy1977a},
because it is practically the only multidimensional Lagrangian technique. The Eulerian
remap process is not necessary in SPH,
so every particle can remember its own evolution history.
Furthermore, by using a kernel with compact support and the tree algorithm \citep{Barnes1986a},
the SPH method is able to achieve a reasonable speed
in finding neighbors and the calculation of self-gravity.

In spite of its various advantages, SPH has some problems.
The long standing one may be the side effect of the artificial viscosity.
The artificial viscosity implemented in the standard SPH \citep{Monaghan1992a}
is very effective to capture shockwaves
and to prevent the particle penetration in the complicated fluid motion,
but has a side effect in a velocity shear.
The problem is that the bulk and shear components cannot be decoupled
in the artificial viscosity of SPH, and the exact contribution of the shear
viscosity in the numerical calculation is hard to manage.
As a result, SPH has difficulty modeling  a Keplerian accretion disk around a compact object or a protostar.
The physical shear viscosity plays an important role in the evolution of an accretion disk
\citep{Shakura1973a,Lynden-Bell1974a,Pringle1981a, Frank2002a},
so accurate handling the physical and numerical
viscosities in a numerical code is very important in
these simulations.

Several approaches have been tried to address this problem.
One of them is the derivation of an empirical viscosity by the modification of the coefficients
appearing in the artificial viscosity \citep{Lubow1991a,Artymowicz1994a}.
Another method is the direct implementation of the physical viscosity in an (inviscid) SPH
code \citep{Flebbe1994a,Takeda1994a,Watkins1996a,Speith1999a,Lanzafame2003a}.
The latter showed good results especially in cataclysmic
variable (hereafter CV) systems typically consisting of a Roche lobe filling low-mass main sequence donor star transferring 
mass to a white dwarf primary star by way of an accretion disk. 

Another approach to solve the side effect of the artificial viscosity
is the use of a Riemann solver instead of the artificial viscosity
\citep{Inutsuka2002a,Cha2003a}. This method is called
GodunovSPH (hereafter GSPH),
because \citet{Godunov1959a} firstly employed the Riemann solver between the adjacent
cells in the finite difference method.
A one--dimensional Riemann solver has been used
on the line joining $i$ and $j$ particles in GSPH,
and provides the time-independent pressure and velocity
at the contact discontinuity as a result. The pressure and velocity
are used in the momentum and energy equations
to update the velocity and the specific internal energy of the particle in the time domain.
The Riemann solver generates a small but sufficient dissipation to describe
shockwaves. \citet{Cha2003a} proved that the Riemann solver can capture high
Mach number shocks, and 
effectively prevents the particle penetration even in a head--on collision of fluids
without the addition of any extra numerical dissipation, such as artificial viscosity.

Another problem of SPH is the numerical surface tension 
\citep{Agertz2007a}.
The momentum equation of standard SPH shows an unphysical force across
the density contrast even in pressure equilibrium, and the unphysical force
acts like a surface tension on the contact discontinuity \citep{Cha2010a}.
The numerical surface tension suppresses the initial perturbation effectively,
so any hydrodynamic instability, such as the Kelvin-Helmholtz instability
cannot grow in the standard SPH simulation. \citet{Price2008a} suggested
artificial conduction for the thermal communications of adjacent particles
to reduce the unphysical force, and showed that standard SPH employing
the artificial conduction can
describe the Kelvin-Helmholtz instability across a density contrast.

A different approach to emphasize the consistency of a particle code
has been suggested by \citet{Inutsuka2002a}. 
He suggests mutually smoothed equations of motion and
energy in GSPH. The mutually smoothed equations consider
the host and neighbor particles as a smoothed one, while  standard SPH 
or the method of \citet{Cha2003a}\footnote{There is an important difference between
\citet{Inutsuka2002a} and \citet{Cha2003a}.
\citet{Inutsuka2002a} uses the mutually smoothed motion and energy equations,
while \citet{Cha2003a} uses the same equations as the
standard SPH. Specifically,
\citet{Inutsuka2002a} considers both of the host and neighbor particles as smoothed, while \citet{Cha2003a} considers only the host particle as
smoothed.
The neighbor particles in the standard SPH and \citet{Cha2003a} are considered
as point masses.
Therefore, we chose that GSPH uses the method of
\citet{Inutsuka2002a}. In order to distinguish the two methods,
\citet{Cha2003a} is called GPH sometimes. If the kernel function of neighbor particles
is assumed to be a delta function, then GSPH is reduced to GPH.}
considers the neighbor particles as point masses.
The mutually smoothed momentum equation is able to get the $1^\text{st}$--order
consistency, and does not show any numerical surface tension \citep{Cha2010a}.
Eventually, GSPH can describe the Kelvin--Helmholtz instability
across the density contrast
without any numerical dissipation term of energy such as the artificial conduction.
Fuerthermore, GSPH can hold magnetohydrodynamics with riemann solvers of a
magnetized fluid \citep{Iwasaki2011a,Tsukamoto2013a,Tsukamoto2015a}

We have developed a parallel GSPH code with the 
Message Passing Interface (MPI\footnote{For the standard
documents of MPI, refer to www.mpi-forum.org/docs}).
With our parallel GSPH code, we are working to simulate the dynamical evolution of a CV accretion disk.
CVs provide a useful testbed for exploring astrophysical viscosity.
They are physically compact and
hence vary on timescales short enough that hundreds to thousands of orbital cycles of time series data can be collected in a
single season from a variety of systems \citep{Wood2007a,Wood2009a}.
The substantial and growing database of
observations provide extremely useful reality check for the theoretical calculations.  While
we are still some years away from fully-physical radiation-magnetohydrodynamical models of
accretion disks, advances in computational hardware and software continue to close that gap.
Without a doubt the question of the nature of astrophysical viscosity is central to the time
evolution of astrophysical accretion disks.

As the first step of this plan, we implemented several numerical routines to handle
the shear viscosity stress tensor in the momentum and energy equations of GSPH.
Furthermore, the implementation has been tested in the context of the Couette flow
\citep{Couette1890a}.
By  comparison of the numerical results with
the analytic solution, we can demonstrate that our implementation works correctly.

In the practical calculation, not only 
the viscous acceleration, but the hydrodynamics acceleration should be calculated
simultaneously. We have performed the full hydrodynamics simulations with
the shear viscosity calculation in Couette flow as well.
All results show the expected result
based on the analytic solution of the genuine Couette flow. 

The kinematic viscosity of a fluid
is not a constant in general. In order to manage the individual kinematic viscosity,
a smoothed dynamic viscosity has been suggested.
We have confirmed that the smoothed dynamic viscosity is able to describe
the viscous acceleration when the particles have a different kinematic viscosity.

We have performed the simulations of the Couette flow with a density contrast to
see the difference between the standard SPH and GSPH.
GSPH shows a less dispersive result, but the difference is not meaningful.

The Couette flow is good to check the correctness of the simulations with viscosity.
However, we need a more realistic test problem to check our implementation.
The evolution of the viscous ring around a central gravity \citep{Lynden-Bell1974a}
has been performed as another test. The results also show an agreement to the
analytic solution.

Sections \ref{sec:gsph} and \ref{sec:implementation} briefly announce the equations of GSPH
and the implementation of the shear viscosity, respectively. Sections \ref{sec:test} and
\ref{sec:viscousring} contain
the simulations of the plane Couette flow and the viscous ring evolution, respectively.
Finally, section \ref{sec:conclusion} presents the conclusion and discussion.

\section{Fundamentals of GSPH}
\label{sec:gsph}
The Lagrangian hydrodynamics of the inviscid fluid is described by
\begin{align}
\label{eq:cintinuity}
\dot \rho & = -\rho\boldsymbol\nabla\cdot \boldsymbol v,\\
\label{eq:momentum}
\dot{\boldsymbol v} & =  -\frac{\boldsymbol\nabla P}{\rho},\\
\label{eq:total_energy}
\dot e & = -\frac{1}{\rho}\boldsymbol\nabla\cdot(P\boldsymbol v),
\end{align}
where $e$ is the specific total energy,
and the other variables have their usual meanings.
Here, the dot above physical quantities means the total derivative (i.e., Lagrangian
derivative with respect to time). 
Instead of the total energy equation, the specific internal energy, $u$ has been
used more frequently in the particle codes, and
the specific internal energy equation becomes
\begin{align}
\label{eq:internal_energy}
\dot u = -\frac P\rho \boldsymbol\nabla\cdot\boldsymbol v.
\end{align}
The equation of state,
\begin{align}
P=(\gamma-1)\rho u
\end{align}
is needed to close the system. Here, $\gamma$ is the specific heat ratio.

With the kernel convolution or the Euler-Lagrange equation,
Eqs. \eqref{eq:momentum} and \eqref{eq:internal_energy} are converted to
the momentum and energy equations of GSPH, respectively \citep{Inutsuka2002a},
\begin{align}
\dot{\boldsymbol v}_i 
\nonumber = &-\sum_j m_j P^*\cdot\\
\label{eq:gsph_mom}
&\left[V^2_{i,j}(h_i)\boldsymbol\nabla_i W_{ij}(\sqrt{2}h_i) + V^2_{i,j}(h_j)\boldsymbol\nabla_i W_{ij}(\sqrt{2}h_j)\right],
\end{align}

\begin{align}
\nonumber \dot{u}_i 
\nonumber= &-\sum_j m_j P^*\left(\boldsymbol v^*-\dot{\boldsymbol x}_i\right)\cdot\\
\label{eq:gsph_ene}
&\left[V^2_{i,j}(h_i)\boldsymbol\nabla_i W_{ij}(\sqrt{2}h_i)
+ V^2_{i,j}(h_j)\boldsymbol\nabla_i W_{ij}(\sqrt{2}h_j)\right],
\end{align}
where the starred quantities, $P^*$ and $\boldsymbol v^*$ are the results of the Riemann solver.
Note that we have used lowercase Greek characters ($\alpha$, $\beta$, \dots)
as the indices of the direction, and $i$ and $j$ as the particle indices in this paper.
We have used the iterative Riemann solver suggested by \citet{vanLeer1997a} to get the starred
quantities.
GSPH describes shockwaves without any numerical dissipation because of the Riemann solver,
and is free from the side effect of the artificial viscosity.
Here, $\dot{\boldsymbol x}_i$ is the updated velocity of the host particle by
$\Delta t/2$, and $\displaystyle V^2_{i,j}$ is the
integral of the squared particle volume. 
$\displaystyle \boldsymbol\nabla_i W_{ij}(\sqrt{2}h_i)$ in
Eqs \eqref{eq:gsph_mom} and \eqref{eq:gsph_ene} is the gradient of kernel function,
given by
\begin{align}
\frac{\partial}{\partial\boldsymbol x_i} W(\boldsymbol x_i - \boldsymbol x_j, \sqrt{2}h_i).
\end{align}
All details of
Eqs. \eqref{eq:gsph_mom} and \eqref{eq:gsph_ene} are described
in Eqs. (61) -- (64) of \citet{Inutsuka2002a}.


\section[]{Implementation of the physical viscosity}
\label{sec:implementation}

\subsection{Viscous stress tensor}
The viscous stress tensor, $\mathbf T$ acting on a fluid element is described by
\begin{align}
\mathbf T = \eta_{\nu}\boldsymbol\sigma + \zeta_\nu\boldsymbol\epsilon,
\end{align}
where $\boldsymbol\sigma$ and $\boldsymbol\epsilon$ are the shear
and bulk viscosity stress tensors, given by
\begin{align}
\label{eq:vtensor}
\sigma_{\alpha\beta} &= \frac{\partial v_\alpha}{\partial x_\beta}
+ \frac{\partial v_\beta}{\partial x_\alpha}
- \frac{2}{3}\delta_{\alpha\beta}\frac{\partial v_\gamma}{\partial x_\gamma},\\
\epsilon_{\alpha\beta} &= \delta_{\alpha\beta} \frac{\partial v_\gamma}{\partial x_\gamma},
\end{align}
component--wise, respectively. Here, $\displaystyle \delta_{\alpha\beta}$ is the Kronecker delta.
Furthermore,
$\eta_\nu$ and $\zeta_\nu$ are the dynamic viscosities of the shear and
bulk components, respectively.
We will omit the bulk viscosity because
the idealized monoatomic gas has been assumed in the simulations.

The kinematic shear viscosity, $\nu$ has a relation
with the dynamic shear viscosity $\eta_\nu$,
\begin{align}
\nu = \frac{\eta_\nu}{\rho}.
\end{align}

\subsection{Acceleration and energy dissipation due to the viscosity}
The acceleration, $\displaystyle\boldsymbol a_\nu$ due to the viscosity is given by
\begin{align}
\boldsymbol a_\nu = \frac{1}{\rho}\boldsymbol\nabla\cdot\mathbf T =
\frac{1}{\rho}\boldsymbol\nabla\cdot\left(\rho\nu\boldsymbol\sigma\right).
\end{align}
The viscosity street tensor contains the gradient of velocity already, so the viscous
acceleration has the second derivative of velocity components. It is well known
that the simple treatment of the second derivative by the standard SPH formulation
is very sensitive to the particle disorder, and therefore too noisy.
It happens not only in the viscous fluid calculation, but in any kind of
simulation including the second derivative, e.g., radiative transfer with the diffusion
approximation \citep{Viau2006a}.
Several previous lines of research have been explored to solve this problem,
and most of them can be categorized into three groups.
\citet{Fatehi2009a} presented a nice review and comparison on this subject.
We will summarize \citet{Fatehi2009a} below for the convenience of the reader.

The first method is the double summation
\citep{Flebbe1994a,Watkins1996a,Speith1999a,Lanzafame2003a}.
Perhaps, it is the most widely
used method in the physical viscosity implementation of SPH. The second derivative of a
physical quantity, $q$ is given by
\begin{align}
\nonumber
\frac{\partial q}{\partial x_\beta}&\left(\kappa\frac{\partial q}{\partial x_\alpha}\right)_i=\\
\label{eq:double_derivative}
&\sum_j\frac{m_j}{\rho_j}
\left(\left<\kappa\frac{\partial q}{\partial x_\alpha}\right>_j
-\left<\kappa\frac{\partial q}{\partial x_\alpha}\right>_i\right)
\frac{\partial W_{ij}}{\partial x_\beta},
\end{align}
where
\begin{align}
\label{eq:sigle_derivative}
\left<\kappa\frac{\partial q}{\partial x_\alpha}\right>_i = \kappa_i\sum_j\frac{m_j}{\rho_j}
\left(q_j - q_i\right)
\frac{\partial W_{ij}}{\partial x_\alpha},
\end{align}
and $\kappa$ is a coefficient of the physical process, e.g., viscosity.
Eq. \eqref{eq:sigle_derivative}
uses a symmetric form of $i-j$ particles, but a non--symmetric form,
\begin{align}
\label{eq:sigle_derivative2}
\left<\kappa\frac{\partial q}{\partial x_\alpha}\right>_i = 
\kappa_i\sum_j\frac{m_j}{\rho_j}q_j
\frac{\partial W_{ij}}{\partial x_\alpha}
\end{align}
can be considered as well. We have used he symmetric form
(Eq. \eqref{eq:sigle_derivative}) all the test calculations below.

In the double summation method,
every particle should calculate the derivative with its own neighbors by
Eq. \eqref{eq:sigle_derivative} firstly.
Then, the second derivative of the
quantity is calculated by Eq. \eqref{eq:double_derivative}.
The essence of the double summation method is to use the information of
the ``neighbors of a neighbor.'' With Eqs. \eqref{eq:double_derivative} and
\eqref{eq:sigle_derivative}, particle $i$ can use not only the information
of particle $j$, but also the information of the neighbors of particle $j$.
Another good property of the double
summation method is that the boundary particles can be managed easily.
Obviously, this method has a fatal disadvantage in the calculation
speed. However, it is inevitable to consider the information the neighbors of a neighbor. 
We have used this double summation method in our implementation.

Another widely used method is the difference scheme
\citep{Brookshaw1985a,Cleary1999a,Monaghan2005b,Viau2006a}.
It uses the single summation to calculate the second derivative,
so should be more efficient in calculation speed.
Furthermore, this method shows a excellent result when particles have
a different $\kappa$, and the contrast of them is very steep
\citep{Cleary1999a,Viau2006a}.
In contrast to the difference scheme, The double summation method has 
difficulty when the difference between $\kappa_i$ and $\kappa_j$ is large.
So we will suggest a smoothed dynamic viscosity,
and show that it is effective in the practical calculations in Sec. \ref{sec:muss}.

The last method is to use the second derivative of the kernel function
\citep{Chaniotis2002a}. We do not consider this method here.

We have employed the double summation method in the viscosity implementation.
First of all, the viscous stress tensor $\boldsymbol\sigma$ has been evaluated
by Eq. \eqref{eq:vtensor}, and then the viscous acceleration is calculated
by the GSPH momentum equation,
\begin{align}
\dot{\boldsymbol v}_{\nu,i}
\nonumber = -\sum_j m_j &\left[\eta_{\nu,i}V^2_{i,j}(h_i)\boldsymbol\sigma_i\cdot\boldsymbol\nabla_i W_{ij}(\sqrt{2}h_i)\right.\\
\label{eq:gsph_avis}
+ &\left. \eta_{\nu,j} V^2_{i,j}(h_j)\boldsymbol\sigma_j\cdot\boldsymbol\nabla_i W_{ij}(\sqrt{2}h_j)\right],
\end{align}
where
\begin{align}
\label{eq:double_dot}
\boldsymbol\sigma\cdot\boldsymbol\nabla_i W_{ij} = \sigma_{\alpha\beta}\frac{\partial W_{ij}}{\partial x_{\beta,i}}\boldsymbol{\hat e}_\alpha.
\end{align}
Here, $\boldsymbol{\hat e}_\alpha$ is the unit vector in $\alpha$-direction.

The viscous coupling between the fluid elements is a typical dissipative mechanism,
so, the effect of the viscosity should be considered
in the energy equation. The equation of the total specific energy
with the viscosity dissipative term becomes
\begin{align}
\dot e  = \dot u + \boldsymbol v\cdot\dot{\boldsymbol v} = -\frac{1}{\rho}\boldsymbol\nabla\cdot
(p\boldsymbol v-\boldsymbol v\cdot\mathbf T).
\end{align}
By comparison with Eq. \eqref{eq:total_energy},
the pure contribution of the viscosity on the internal energy is given by
\begin{align}
\dot u_\nu = \frac{1}{\rho}\boldsymbol\nabla\cdot(\boldsymbol v\cdot\mathbf T).
\end{align}
With the similar way for the momentum equation, the viscous energy equation becomes,
\begin{align}
\dot u_{\nu,i}
\nonumber = -&\sum_j m_j \left[\eta_{\nu,i}V^2_{i,j}(h_i)(\boldsymbol v_i\cdot\boldsymbol\sigma_i)\cdot\boldsymbol\nabla_i W_{ij}(\sqrt{2}h_i)\right.\\
\label{eq:gsph_evis}
+ &\left. \eta_{\nu,j} V^2_{i,j}(h_j)(\boldsymbol v_j\cdot\boldsymbol\sigma_j)\cdot\boldsymbol\nabla_i W_{ij}(\sqrt{2}h_j)\right],
\end{align}
where
\begin{align}
\label{eq:triple_dot}
(\boldsymbol v\cdot\boldsymbol\sigma)\cdot\boldsymbol\nabla_i W_{ij} = v_\alpha\sigma_{\alpha\beta}\frac{\partial W_{ij}}{\partial x_{\beta,i}}.
\end{align}
There is no ambiguity in Eqs. \eqref{eq:double_dot} and \eqref{eq:triple_dot}
because $\boldsymbol\sigma$ is a symmetric tensor.
A full description on the viscous energy dissipation can be found in \citet{Flebbe1994a}.

The viscous acceleration and energy dissipation are
added to Eqs. \eqref{eq:gsph_mom} and \eqref{eq:gsph_ene} to make
the final acceleration and internal energy increment by the operator splitting method,
respectively.

\section[]{Plane Couette flow}
\label{sec:test}
The plane Couette flow (hereafter Couette flow) \citep{Couette1890a}
describes the motion of a viscous flow between two moving parallel plates.
One plate is at the top, and the other is at the bottom of the fluid.
The plates at the top is moving with a constant velocity along the direction of the fluid
extension, and the bottom one is moving by a different velocity or at rest. With the
non-slip boundary condition between the plates and the fluid, the fluid gains momentum
from the plate movement due to the viscous coupling. Although it is the simplest shear flow
which follows Newton's law of viscosity, actual experiments have been rarely done
because of practical difficulties  \citep{Kitoh2005a}.

We assume a two--dimensional situation. The fluid is at rest initially,
and the plate at the top of the fluid ($y=L$)
is moving right with a constant velocity, $v_o$, while the plate at the fluid bottom ($y=0$)
is at rest. The Couette flow under this circumstance is described by
\begin{align}
\frac{\partial}{\partial t}v_x (y,t) = \nu \frac{\partial^2}{\partial y^2} v_x(y,t),
\end{align}
where $v_x$ is the $x$--component of the fluid velocity. Here, the kinematic viscosity
has been assumed as a constant all over the fluid.
It is the simplest diffusion equation with an inhomogeneous boundary condition,
\begin{align}
\label{eq:bdcond}
\begin{cases}
v_x(L,t) & = v_o\\
v_x(0,t) & = 0 
\end{cases}
\end{align}
and the initial velocity profile,
\begin{align}
v_x(y,0) = 0,
\end{align}
where $L$ is the calculation domain size.
The analytic solution is to be found easily by the separation of variables, and is given by
\begin{align}
\label{eq:ana_sol}
v_x(y,t) = v_{o}Ly+ \sum_n(-1)^n\frac{2v_{o}}{n\pi}e^{-\left(\frac{n\pi}{L}\right)^2 \nu t}\sin\left(\frac{n\pi}{L} y\right).
\end{align}

\subsection{Problem setup}
We have used two--dimensional GSPH in the Couette flow simulations, and
all physical quantities in the code are dimensionless. A unity box ($1\times 1$)
with $100\times 100$ grids in the $xy$--plane has been used in the calculation.
Initially, the particles are uniformly positioned at the center of the grids
to minimize the numerical noise.

The moving blocks are treated as the boundary zone in the simulations,
so they are beyond the calculation domain.
The thickness of the blocks in $y$--direction is 0.2, and is enough to cover the several times of
the smoothing length of the fluid particles.

The upper moving block should have a constant velocity, $v_o$,
and the bottom is at rest in the Couette problem. However,
the moving blocks are considered as a boundary zone in the simulations, so
the upper and lower blocks should have a variable velocity. It is because 
the the non--slip boundary condition has been employed (i.e., extremely strong viscous coupling)
in the tests.
The value of $v_o$ is set to $0.1$ in the simulation, but the actual
velocity of the moving blocks has been determined
by the extrapolation of the analytic solution (Eq. \eqref{eq:ana_sol}) to the boundary zone
at every evolution time.

Another practical problem is the startup velocity of the fluid.
A sudden movement of the top block with respect to the fluid
may cause a slip, and a noise in the velocity profile appears at the very early
stage. The noise won't decay even in the later evolution.
So, we have set an the initial time, $t_o$,
\begin{align}
t_o = \frac{1}{200}t_\nu,
\end{align}
and determine a corresponding velocity of the fluid at $t=t_o$ by
Eq. \eqref{eq:ana_sol} as the startup velocity.
Here, $\displaystyle t_\nu$ is the viscous timescale, given by
\begin{align}
t_\nu = \frac{L^2}{\nu}.
\end{align}
For the startup velocity of the fluid, one can use
\begin{align}
\label{eq:block_vel}
v_{x}(y,t_o) = v_o \text{erfc}\left(\frac{L-y}{2\sqrt{\nu t_o}}\right)
\end{align}
instead of Eq. \eqref{eq:ana_sol}. 
Here, erfc($x$) is the complementary error function.
Eq. \eqref{eq:block_vel}  is more convenient  in the simulations.

Two cases of simulations have been performed. One is the genuine Couette flow.
The fluid particles are moved by the viscosity
acceleration only in this case. Therefore, any hydrodynamic acceleration or energy calculation have
not been involved in this case. However, note that neighbor finding, the density update
and the individual smoothing length determination based on the updated density
have been calculated. So, one can see the numerical dispersion in this case if it exists.

The other case simulated here is the full hydrodynamics calculation with the viscosity.
In this case, hydrodynamics acceleration and energy calculation
 by Eqs. \eqref{eq:gsph_mom} and \eqref{eq:gsph_ene} have been added
to the viscous acceleration and energy dissipation.
There is no analytical description in this case, however,
one may expect the almost same result of the former case. It is because
the pressure equilibrium has been assumed, and the main driving force of the fluid motion
is the viscous coupling with the moving plates rather than hydrodynamics.
We expect that the latter may be useful to check the influence of the
numerical dispersion in our GSPH code.

The density of the fluid is set to the unity initially in the whole calculation domain,
and the pressure is set to
\begin{align}
p=\frac\rho\gamma,
\end{align}
where, $\gamma$ is the specific heat ratio, and is set to $5/3$ in the simulations with
hydrodynamics.
This pressure value makes the sound speed unity.

The calculations with hydrodynamics have been performed
with the second-order GSPH code.
Refer to \citet{Inutsuka2002a} and \citet{vanLeer1997a} to see
the details of the second-order scheme.

\subsection{Results of the Couette flow}
We have performed several simulations with various kinematic viscosities, from $10$ to
$1\times 10^{-2}$, and all simulations show the same result at each normalized evolution
time by the viscosity timescale. 
\begin{figure*}
\centering
\includegraphics[width=0.4\textwidth]{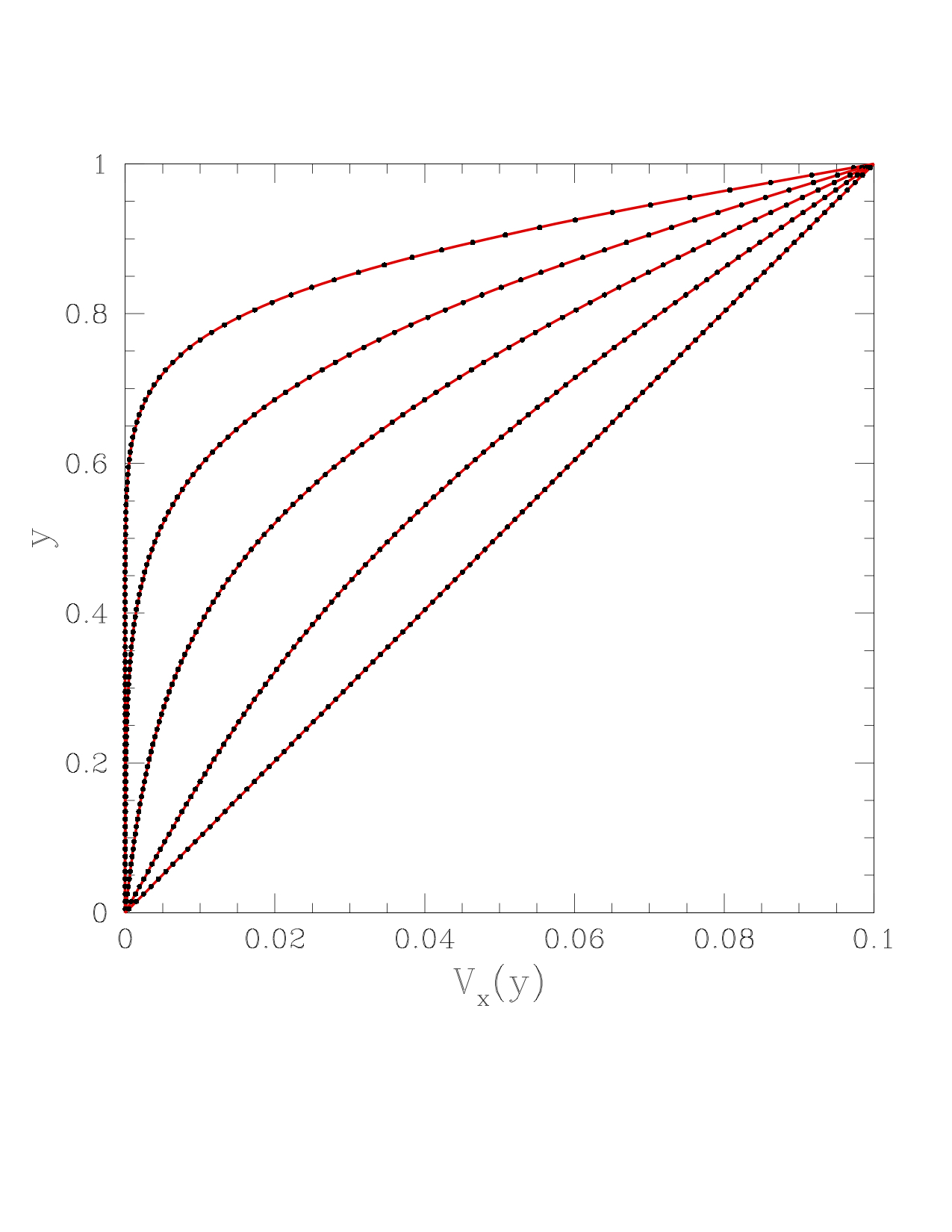}
\includegraphics[width=0.4\textwidth]{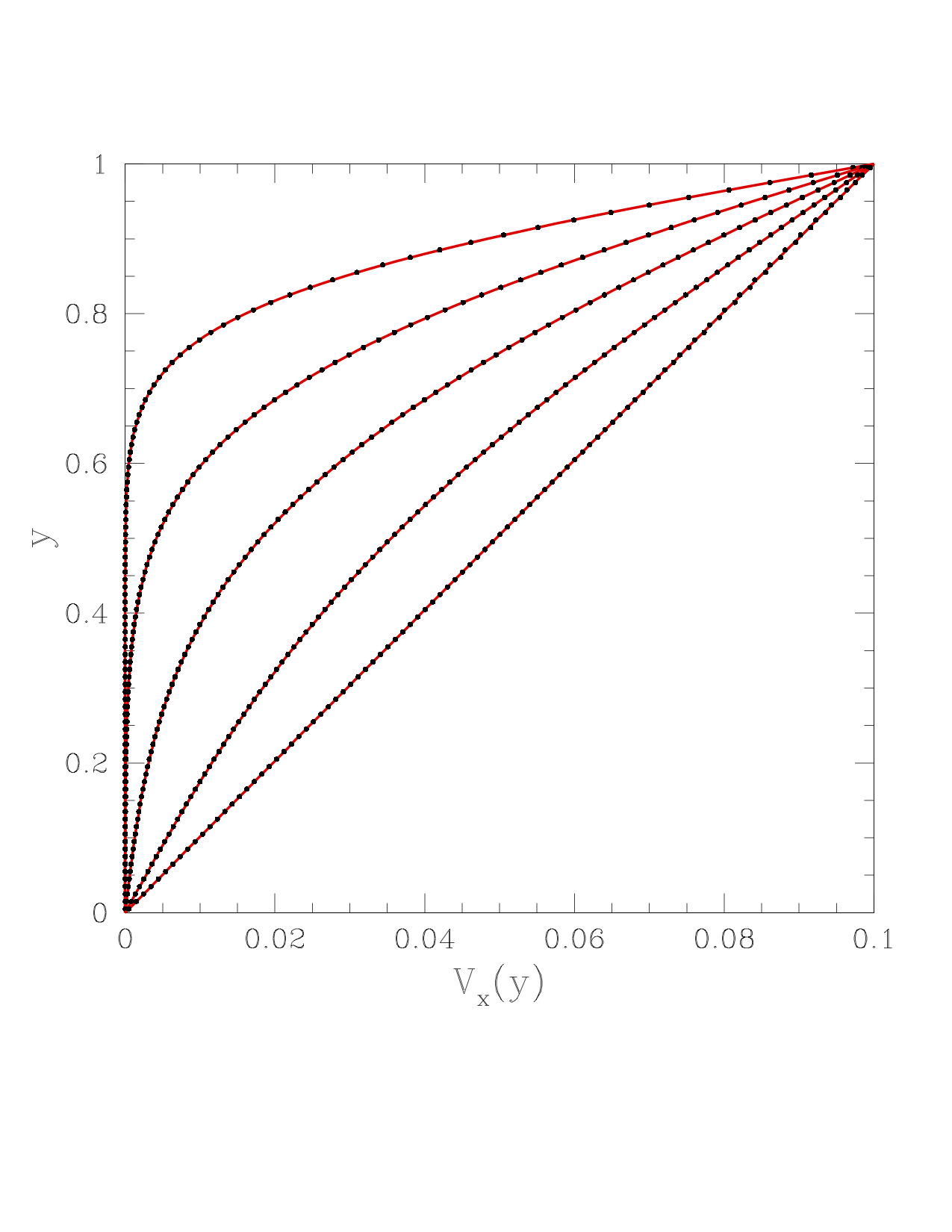}
\caption{Simulations of the Couette flow with $\nu=1.0$ and $0.1$ in left and right
panels, respectively. These are the velocity--position plots of the viscosity--only calculations.
The red solid lines are the analytic solution given by Eq. \eqref{eq:ana_sol} at 
$t=0.1,\,\,0.3,\,\,0.7,\,\,1.5,\,\,\text{and}\,\,5\times10^{-1}t_\nu$ from the left to the right,
and the black dots are the numerical solution at the same time.
Around $\displaystyle \sim 0.5 t_\nu$, the fluid nearly achieved the equilibrium solution.}
\label{fig:normal_nu}
\end{figure*}
Fig. \ref{fig:normal_nu} shows the position-velocity plots
of the simulations of $\nu = 1.0$ and $0.1$ cases without hydrodynamics.
The red solid lines are the analytic solution calculated by Eq. \eqref{eq:ana_sol},
and the black dots are the numerical results.
The numerical and analytic solutions show a good agreement, and
the equilibrium solution,
\begin{align}
v_x(y) = \displaystyle v_{o}Ly
\end{align}
has been almost obtained around $t\simeq0.5t_\nu$.
These are two--dimensional simulations, so one black dot in the figures
represents a group of fluid
particles rather than a single fluid particle.
One can see the suppressed numerical dispersion. Note that these simulations
calculate the viscosity force only, but the other GSPH routines, e.g., finding neighbors,
the density estimation and the smoothing length determination are used in the calculation.

\begin{figure*}
\centering
\includegraphics[width=0.4\textwidth]{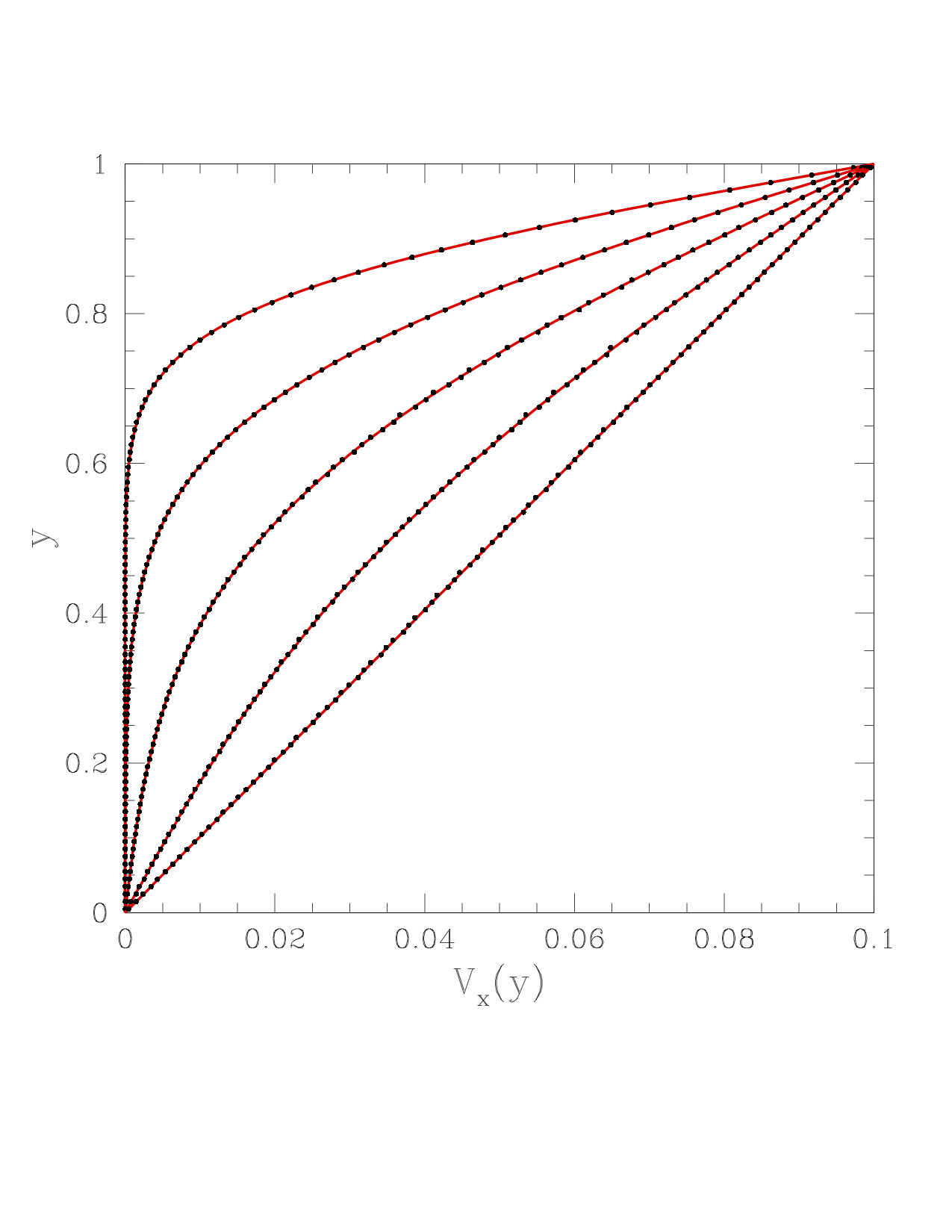}
\includegraphics[width=0.4\textwidth]{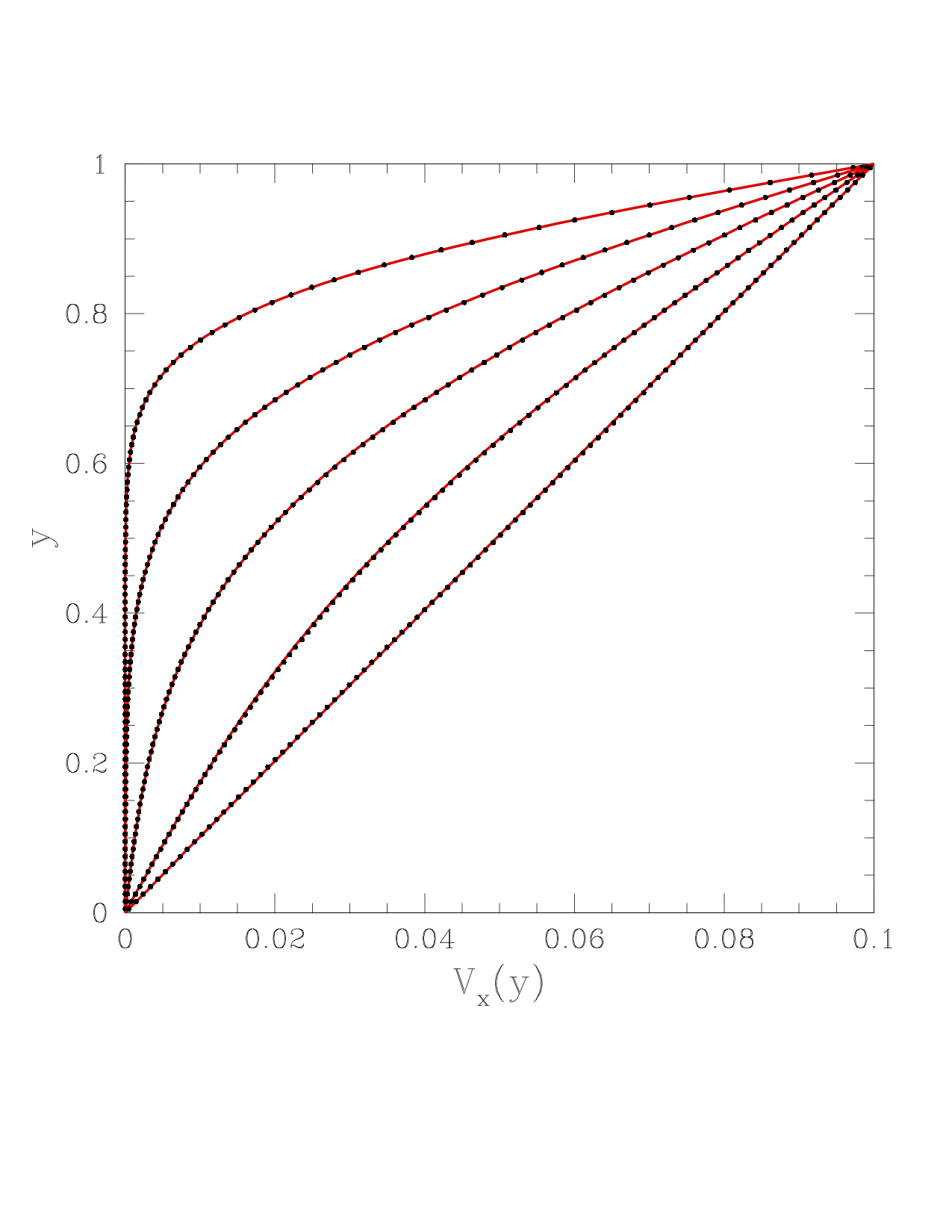}
\caption{This is the same simulations as Fig. \ref{fig:normal_nu},
but hydrodynamics included. The kinematic viscosity and other configurations
of the simulations of left and right
panels are the same as Fig. \ref{fig:normal_nu}.
The contribution of the pressure gradient are negligible, so the results are nearly
the same as the simulations without hydrodynamics.
One can see that the numerical dispersion is suppressed.}
\label{fig:normal_nuhd}
\end{figure*}

The viscosity force calculation by Eq. \eqref{eq:gsph_evis} has been confirmed
successfully by the tests above,
but we have to check the overall performance of our viscous GSPH code.
It is because the hydrodynamic acceleration and adiabatic changes of the internal energy
are also important
in the viscous fluid motion. So, we have performed
the same simulations again with a full steps of hydrodynamics. Fig. \ref{fig:normal_nuhd}
is the results with hydrodynamics.
The kinematic viscosity and other configurations of the
simulations are the same as the viscosity--only simulations.
Although there is no analytic solution in this case,
one may expect nearly the same results as the viscosity--only calculation,
because the contribution of the pressure gradient is negligible.
Essentially, there is no significant difference in Figs. \ref{fig:normal_nu} and \ref{fig:normal_nuhd},
as expected.

\subsection{High Reynolds number simulations}
The Reynolds number defined by
\begin{align}
Re = \frac{Lv_o}{\nu}
\end{align}
has been frequently
used as a parameter of the possibility of turbulence. The simulations in the previous
section have the kinematic viscosity range in $\displaystyle 10\sim1\times10^{-2}$,
and the corresponding $Re$ values are in the range $10^{-2}\sim10$. No dispersion  
appears in this $Re$ range.

We have performed some simulations with a higher $Re$ number to define the possibility of 
the occurrence of turbulence. The configuration of the simulations
has been changed in these tests.
Specifically, the two blocks at the top and bottom of the fluid
are moving in opposite directions with the same speed. There is
an analytic solution to this boundary condition, given by
\begin{align}
\nonumber
v_x(y,t) &= v_{o}\left(\frac{2y}{L}-1\right)\\
\label{eq:ana_sol2}
& + \sum_n(1+(-1)^n)\frac{2v_{o}}{n\pi}e^{-\left(\frac{n\pi}{L}\right)^2 \nu t}\sin\left(\frac{n\pi}{L} y\right).
\end{align}
The system size $L$ in Eq. \eqref{eq:ana_sol2} is set to $1$, and the plate moving velocity,
$v_o$ is set to $0.1$.

\begin{figure*}
\centering
\includegraphics[width=0.4\textwidth]{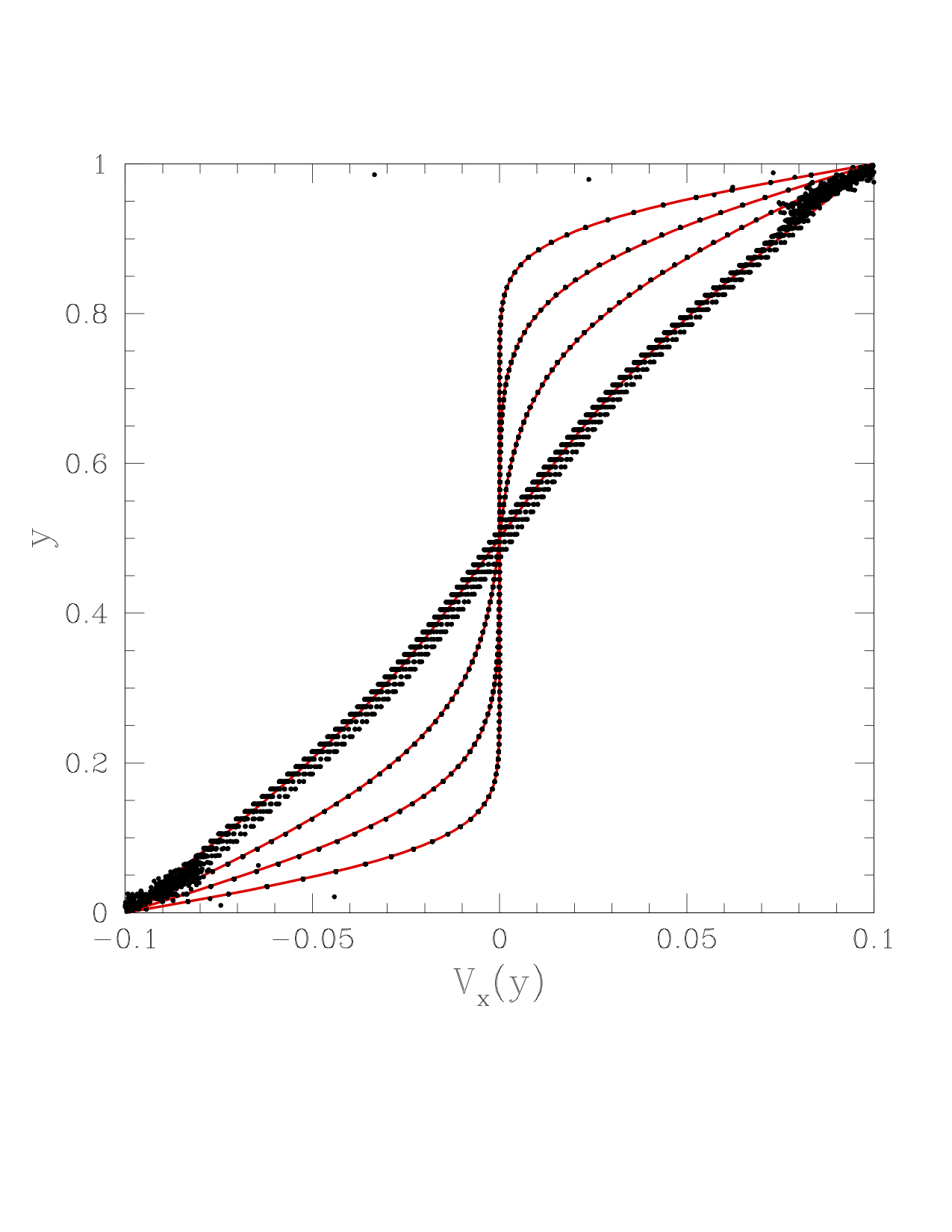}
\includegraphics[width=0.4\textwidth]{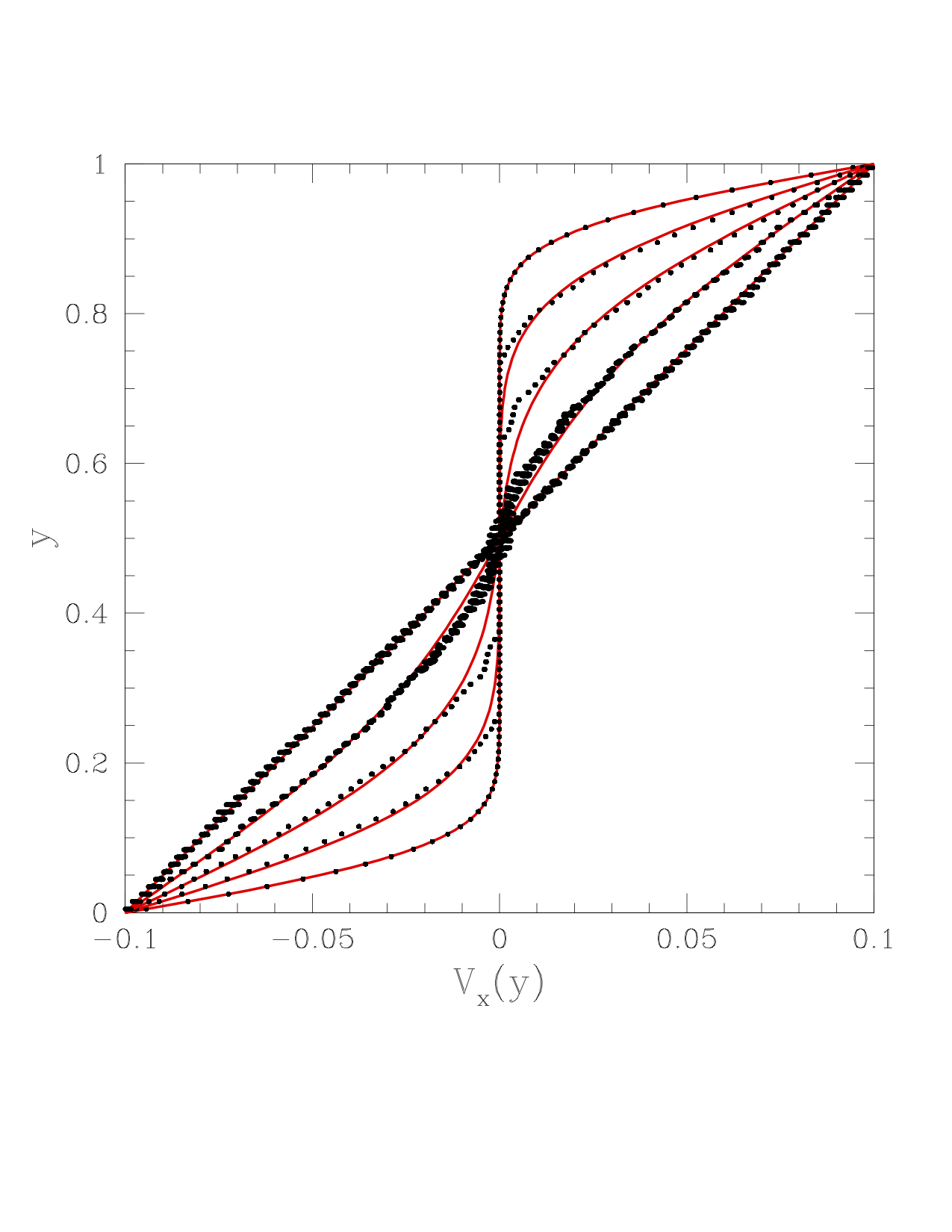}
\caption{The simulations with a higher $Re$ value. The value of $Re$ is set to $100$, and
the corresponding kinematic viscosity is $10^{-3}$. The left and right panels 
are without and with hydrodynamics calculations, respectively. Note that the output times
of the left and right panels are different. It is because the viscosity--only calculation
with a high $Re$ value
becomes unstable in the later time. In the left panel,
$t=0.1,\,\,0.3,\,\,0.7,\,\,\text{and}\,\,2\times10^{-1}t_\nu$, and the right panel,
$t=0.1,\,\,0.3,\,\,0.7,\,\,1.5,\,\,\text{and}\,\,5\times10^{-1}t_\nu$.
They are more dispersive than the smaller $Re$ cases. 
}
\label{fig:small_nu}
\end{figure*}
Fig. \ref{fig:small_nu} shows the results of the high $Re$ simulations. The left and
right panels show the results without and with hydrodynamics, respectively. Note that the output
times are different in the two panels because the result of the viscosity--only calculation
becomes very unstable in the later times. The kinematic viscosity
used in the simulations is $10^{-3}$, and the corresponding $Re$
value becomes 100. The fluid becomes turbulent around $t\sim t_\nu/4$. 
Although it is still controversial, the Couette flow
is expected to be  turbulent around $Re \simeq 325$ \citep{Schneider2010a}.
Our results show the occurrence of turbulence at a smaller $Re$ value more or less.
Another interesting point is that the full hydrodynamics
calculation is more stable (i.e., less dispersive in the final snapshot) than the pure viscosity calculation.

\begin{figure*}
\centering
\includegraphics[width=0.8\textwidth]{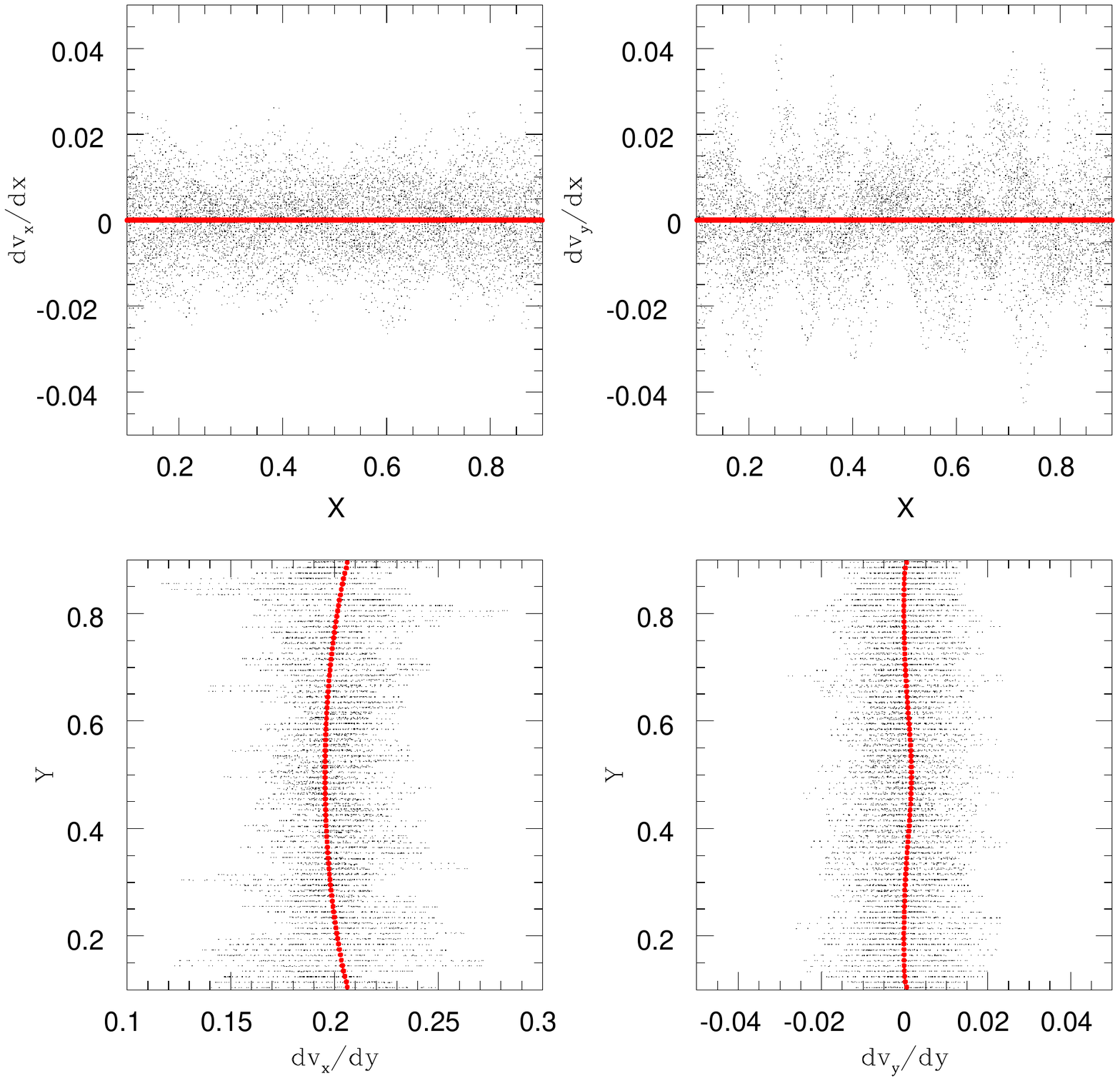}
\caption{The dispersion of velocities in $x$-- and $y$--directions. The red thick
and black thin dots are the results of $\nu=10^{-1}$ and $10^{-3}$ cases, respectively.
The corresponding $Re$ values are $1$ and $100$, respectively.
More dispersive pattern arises in the
high $Re$ case, and the severe dispersion may cause a turbulence. In the low $Re$ case,
the curved feature of $\displaystyle dv_x/dy$ in $y$--direction is due the the imperfect
settling to the equilibrium solution at $t=0.5t_\nu$.}
\label{fig:dis}
\end{figure*}

Fig. \ref{fig:dis} shows the velocity dispersion along $x$-- and $y$--directions. The red big
and black dots are the results of $Re=1$ and $100$, respectively. The both are
the results of the full hydrodynamics calculations.
First of all, the low $Re$ case show very clean profiles in $x$--direction without
any dispersion. The curved features of the low $Re$ case
in $\displaystyle dv_x/dy$ along $y$--direction is due the
imperfect settling to the equilibrium solution. Fig. \ref{fig:dis} is at $t=0.5t_\nu$, and
one may expect a less curved pattern around $t\sim t_\nu$.
One can see a very dispersive pattern in the high $Re$ simulation.
The viscosity coupling between the moving blocks and the fluid acts on
$v_x$ in $y$--direction only, but all velocity components in every direction gain an
acceleration, and the fluid shows the turbulent pattern eventually.

\subsection{A smoothed dynamic viscosity : Implementation and test}
\label{sec:muss}
We have tested our viscosity managing routines with a universal kinematic viscosity for all
particles so far. However, the kinematic viscosity of a particle is not constant over the calculation
domain in general, and is a function of the physical quantities of the fluid
(e.g., density and temperature). We have found that
Eqs. \eqref{eq:gsph_avis}, \eqref{eq:gsph_evis} are not effective when the the host
and neighbor particles have a different kinematic viscosity.
A smoothed dynamic viscosity has been introduced to solve
this problem.
The smoothed dynamic viscosity is determined by
\begin{align}
\nonumber
\bar\eta_{\nu,i} &= \sum_j m_j \int \frac{\eta_{\nu,j}}{\rho(\boldsymbol x)}W(\boldsymbol x - \boldsymbol x_i, h_i)
W(\boldsymbol x - \boldsymbol x_j, h_j)d \boldsymbol x\\
&\simeq \sum_j m_j \nu_j W_{ij}(\sqrt{2}h_{ij}),
\end{align}
where ${\eta_{\nu,i}}$ is approximated by
\begin{align}
{\eta_{\nu,j}} \simeq \rho(\boldsymbol x)\nu_j.
\end{align}
Note that the kinematic viscosity of the individual particle is not a smoothed quantity,
and should be determined in advance by the physical conditions of the particle.

The same momentum and energy equations, Eqs. \eqref{eq:gsph_avis} and \eqref{eq:gsph_evis}
are used, but the dynamic viscosities, $\eta_{\nu,i}$ and $\eta_{\nu,j}$
of the equations should be replaced
by the smoothed dynamic viscosities, $\bar\eta_{\nu,i}$ and $\bar\eta_{\nu,j}$.

Two layers of a viscous flow have been considered as a test problem.
The fluid is assumed to be extended infinitely in the $x$-- and $y$--directions, and
has different kinematic viscosities in the upper $(y>0)$ and lower $(y<0)$ layers.
Contrary to the genuine Couette flow, there is no moving solid block in this test,
but the upper layer is moving by a initial velocity, $v_o$ to the right, while
the lower layer is at rest initially. The viscous coupling between the two layers
will generate an acceleration in the layers. The kinematic viscosity
is a function of $y$--poisition only, so they are given by
\begin{align}
\nu_i = 
\begin{cases}
\nu_U & \text {if }y>0\\
\nu_L & \text {otherwise}
\end{cases}.
\end{align}

There is an analytic solution for this infinitely extended viscous flow
\citep{Carslaw1959a}, and the $x$--component of the fluid velocity is give by
\begin{align}
\label{eq:ana_up}
v_{x,U} &= \frac{v_o\mathcal K_U}{\mathcal K_U+\mathcal K_L}\left[1+\frac{\mathcal K_L}{\mathcal K_U}\text{erf}\left(\frac{y}{2\sqrt{\nu_U t}}\right)\right]
\end{align}
for the upper layer, and
\begin{align}
\label{eq:ana_low}
v_{x,L} &= \frac{v_o\mathcal K_U}{\mathcal K_U+\mathcal K_L}\text{erfc}\left(\frac{\left\vert y\right\vert}{2\sqrt{\nu_L t}}\right)
\end{align}
for the lower layer.
Here, $\displaystyle\mathcal K = \sqrt{\rho\eta_\nu}$,
and erf$(x)$ is the error function. The value of $v_o$ is set to 0.1 in the test.

\begin{figure*}
\centering
\includegraphics[width=0.4\textwidth]{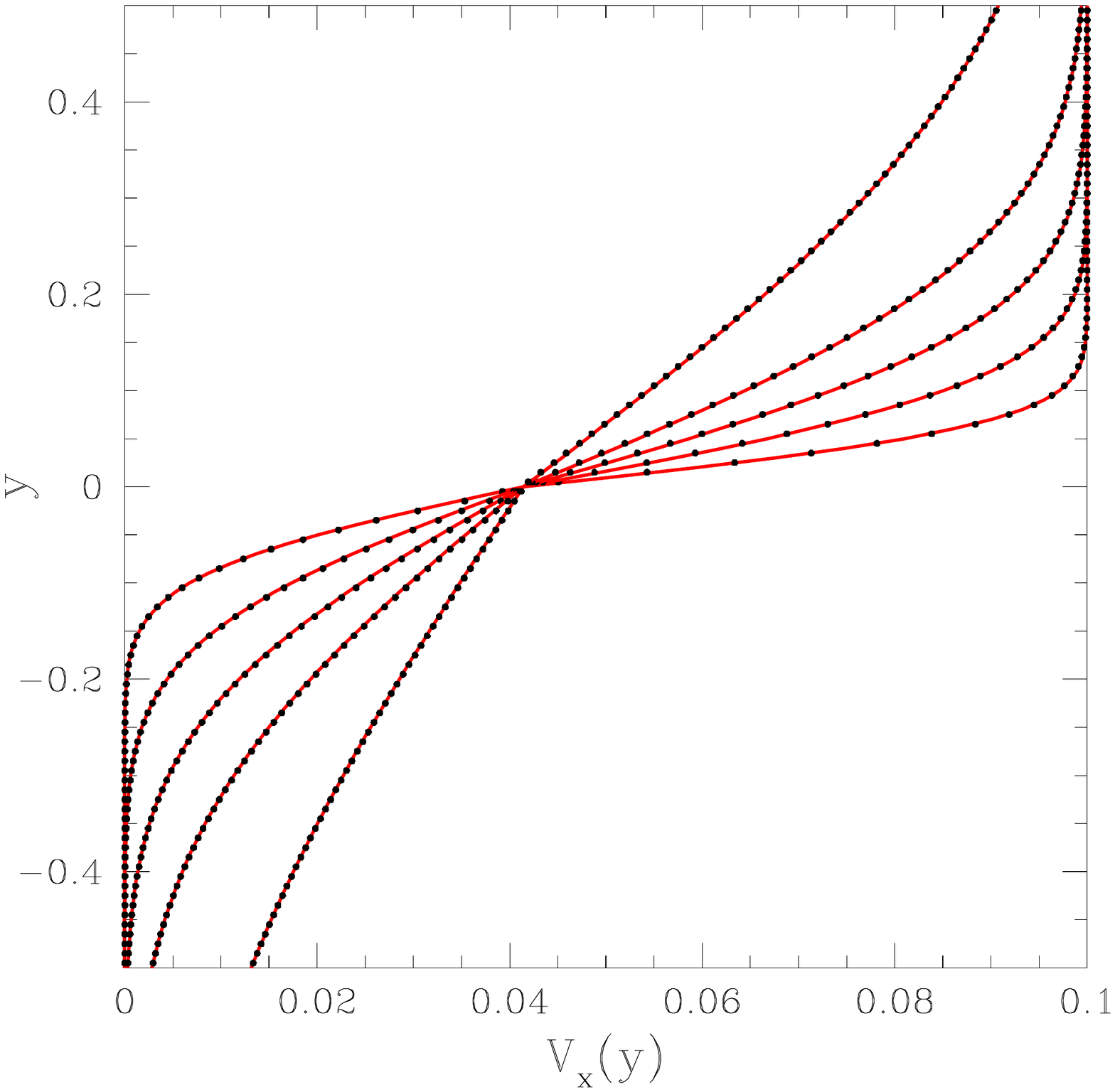}
\includegraphics[width=0.4\textwidth]{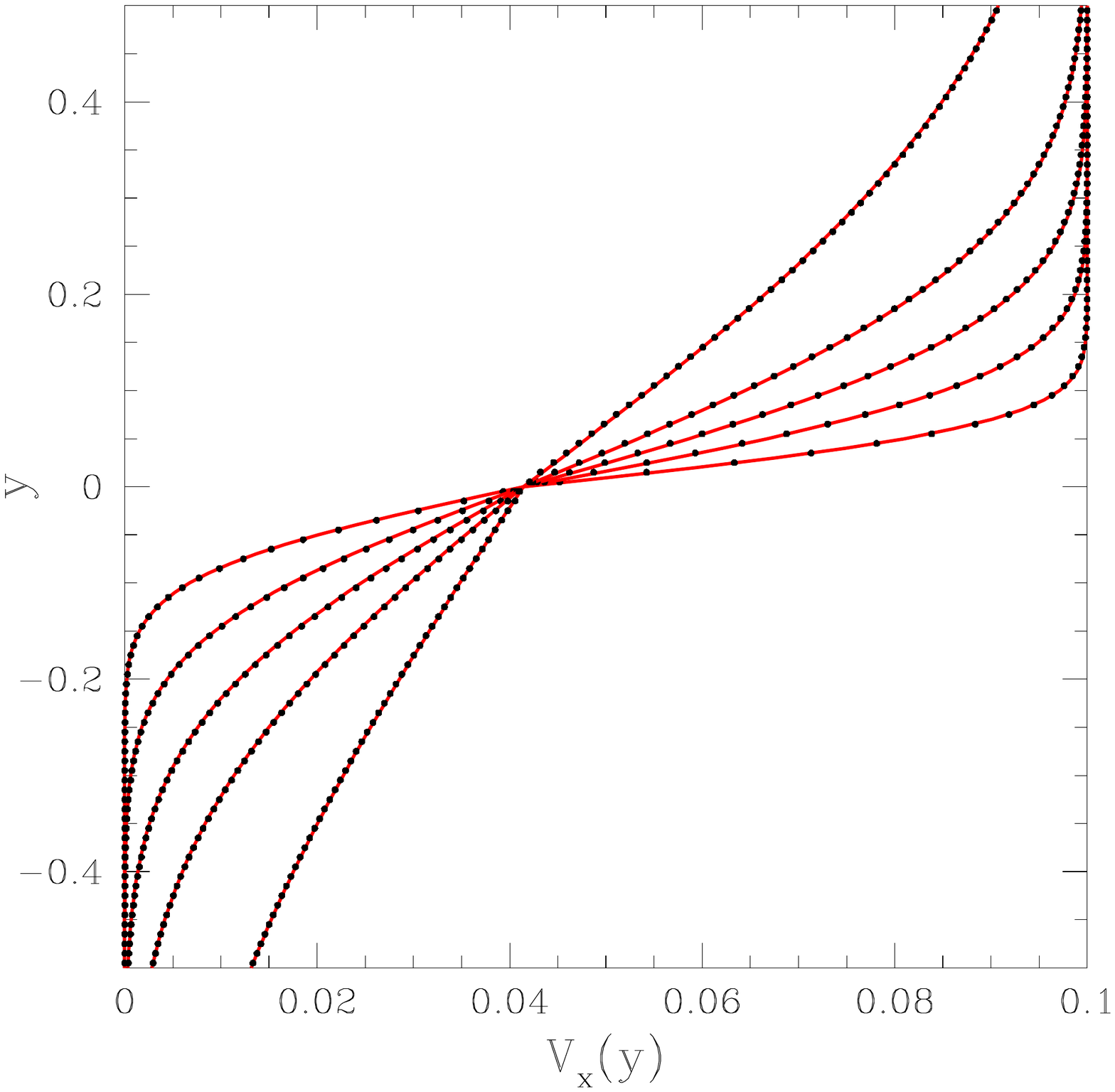}
\caption{Two layers simulation results at
$t=0.1,\,\,0.3,\,\,0.7,\,\,1.5,\,\,\text{and}\,\,5\times10^{-1}t_\nu$. The steeper profile is the later
result. The left and right panels are without and with
hydrodynamics, respectively. The red solid lines and black dots are the analytic and numerical
solutions, respectively. The two solutions agrees with each other well.
The kinematic viscosity of the upper
$(y>0)$and lower $(y<0)$ layers are 0.1 and 0.2, respectively.}
\label{fig:invicid}
\end{figure*}
Fig. \ref{fig:invicid} is the results of the two--layered viscous flow. The kinematic viscosities of the
upper and lower layers are $0.1$ and $0.2$, respectively. The output times of the figure
are scaled by the viscous time scale, defined by
\begin{align}
\label{eq:tnu:diff_vis}
t_\nu = \frac{0.5^2}{\text{max}(\nu_U,\,\,\nu_L)}.
\end{align}
The numerical solution (black dots) coincides with the analytic solution (red solid lines) very well,
and numerical dispersion does not appear at all.

\begin{figure*}
\centering
\includegraphics[width=0.4\textwidth]{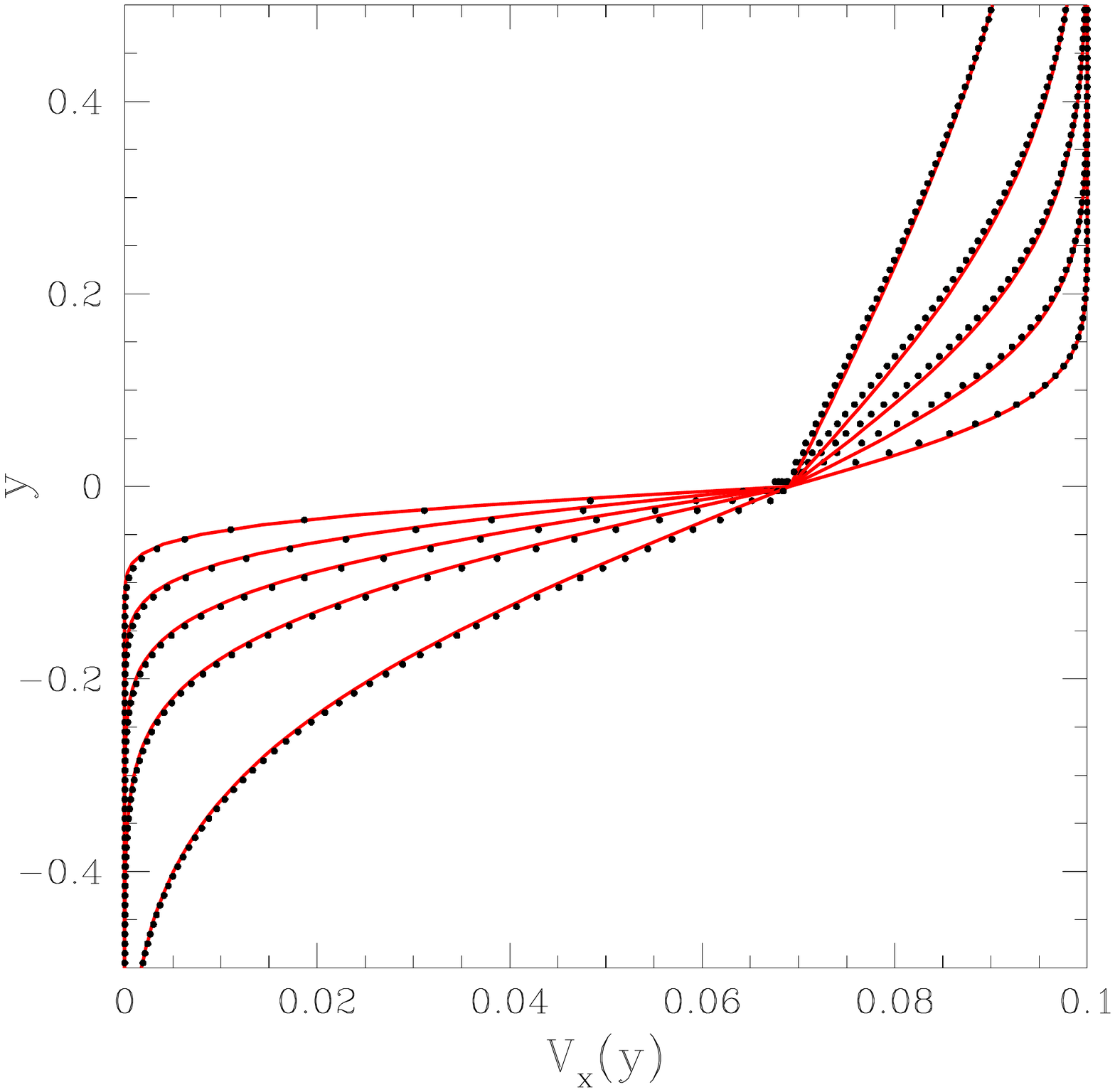}
\includegraphics[width=0.4\textwidth]{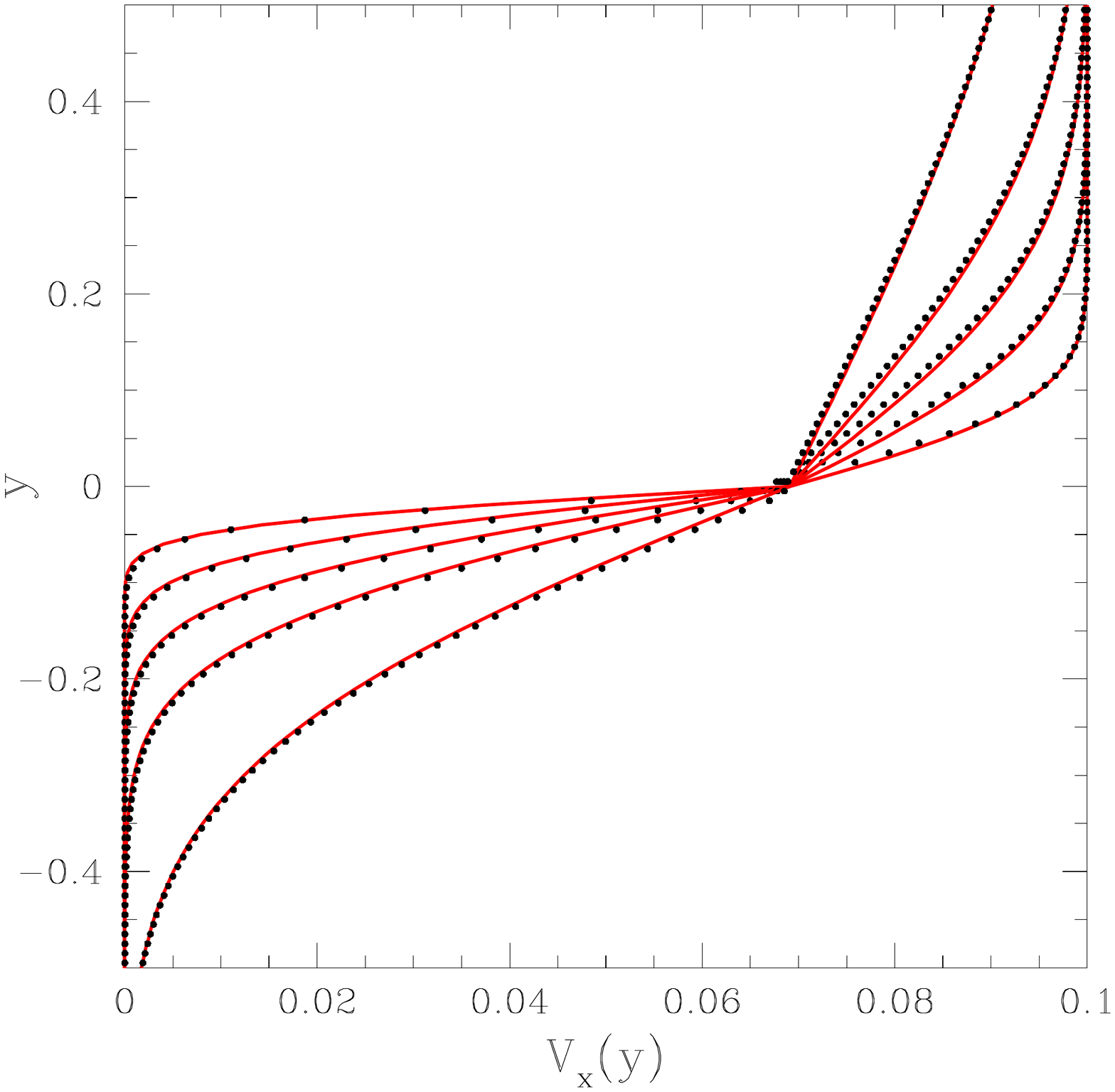}
\caption{Two layers simulations of a higher contrast in the kinematic viscosity.
The kinematic viscosities of the upper and lower layers are 0.5 and 0.1, respectively.
The agreement between the numerical and analytic solutions are still good.
The output times are the same as Fig. \ref{fig:invicid}.}
\label{fig:invicid2}
\end{figure*}
Fig. \ref{fig:invicid2} is the results of the higher contrast in the kinematic viscosities.
The kinematic viscosity of the upper layer is 5 times bigger than the lower layer. Still,
the numerical and analytic solutions show a good agreement.

\subsection{Test with a density contrast : Comparison with the standard SPH}
\label{sec:comp_sph}
Perhaps one interesting subject is the comparison between the standard SPH and GSPH. However,
the Couette flow should not be a good example to show the difference of the two numerical methods,
because there is no hydrodynamics in the Couette flow.
Furthermore, it is hard to find a significant difference in the two methods even in the full hydrodynamics
calculation.
It is obvious because the Couette flow assumes the pressure equilibrium,
so the hydrodynamic force is negligible compare to the viscous one as presented
in Fig. \ref{fig:normal_nuhd}.

Another consideration is necessary to see the difference. So we have performed
the hydrodynamic Couette flow tests with a density contrast.
As we have described in Sec. \ref{sec:intro}, the standard SPH shows the
numerical surface tension across the density contrast, so one may expect to see a difference
in the results of the standard SPH and GSPH.

Essentially, all configurations of the simulation are the same as the two layers simulations
in the previous section except the
density contrast.
The lower layer is denser and more viscous than the upper one by 4 and 2 times, and
the density and viscosity of the lower layer are set to 4 and 0.2, respectively.
Note that the contrast of the dynamic viscosity becomes 8 across the two layers.

Initially, the upper layer
moves to the right by the velocity of 0.1. Without the hydrodynamics, the analytic solution given
in Eqs. \eqref{eq:ana_up} - \eqref{eq:ana_low} can be used by the modified
$\displaystyle\mathcal K_U (= \sqrt{\rho_U\eta_{\nu,U}})$ and
$\displaystyle\mathcal K_L (= \sqrt{\rho_L\eta_{\nu,L}})$. The evolution time of the simulations
are scaled by the viscous time scale, given in Eq. \eqref{eq:tnu:diff_vis}.

\begin{figure*}
\centering
\includegraphics[width=0.4\textwidth]{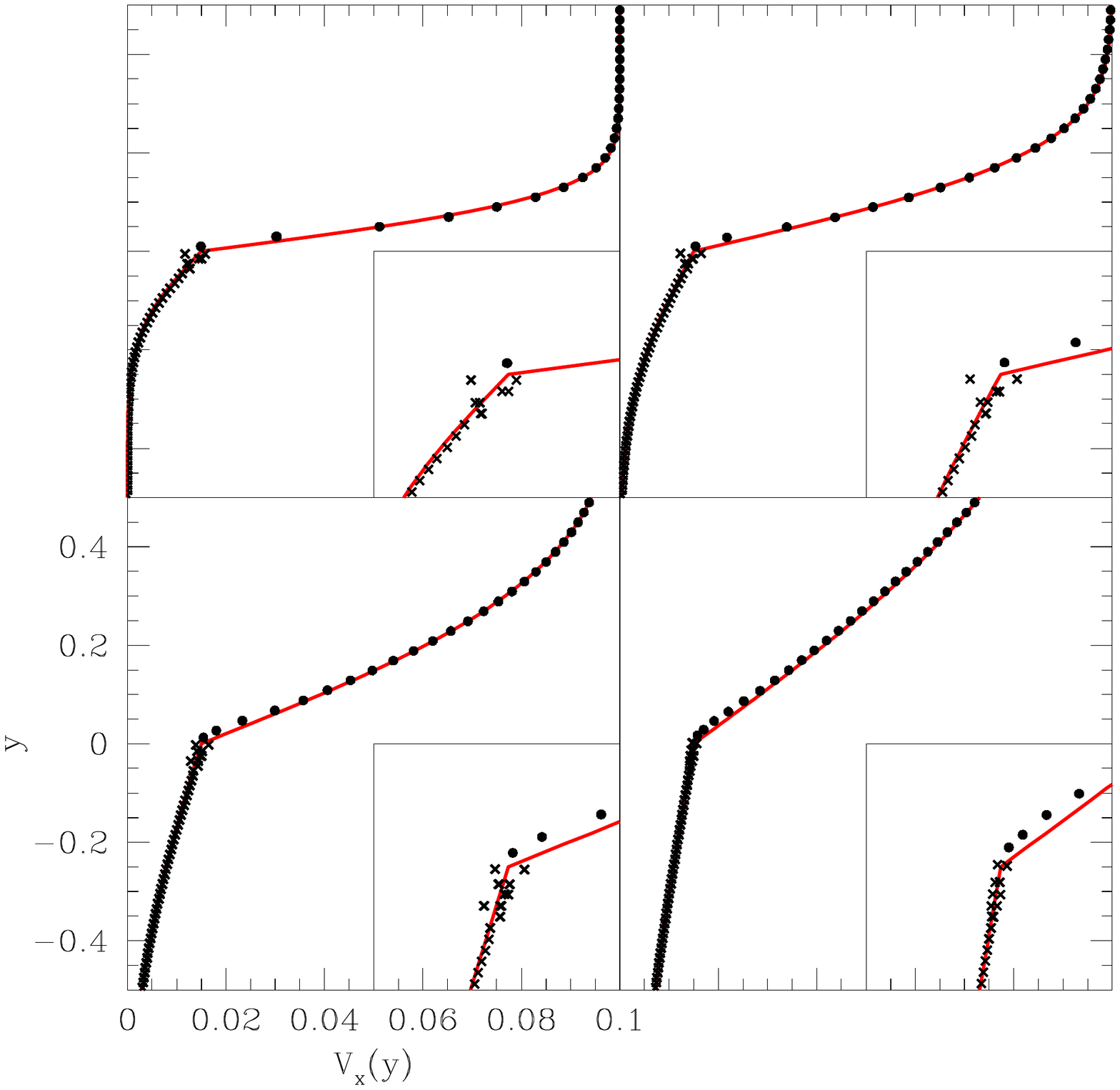}
\includegraphics[width=0.4\textwidth]{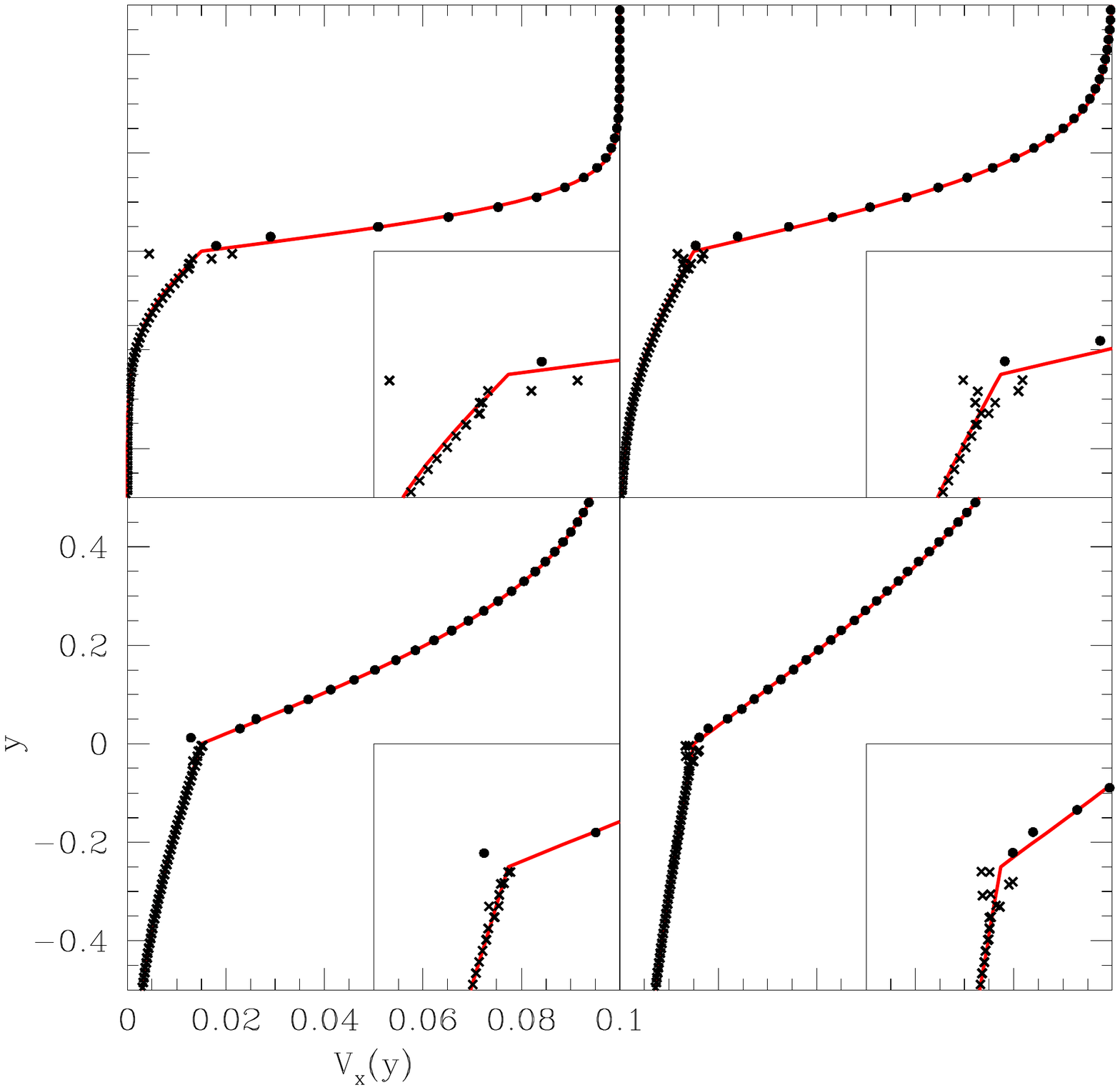}
\caption{Simulations of the Couette flow with a density contrast. Each panel shows the result at
$t=3,\,\,10,\,\,30,\,\,\text{and}\,\,100\times10^{-2}t_\nu$ from the upper left to the lower right.
The left- and right-hand-side are the results of GSPH and the standard SPH, respectively.
The embedded small boxes are the closeup at the boundary of the two different density layers. The red solid
lines are the analytic solution given in Eqs. \eqref{eq:ana_up} - \eqref{eq:ana_low}, and the
solid circles and crosses are the upper and lower fluid, respectively. No significant
difference appears, but the standard SPH simulation shows a more dispersive
pattern at the boundary of the layers. The lower layer is denser and more viscous,
and the density and kinematic viscosity of the lower layer
are set to 4 and 0.2 in the code unit, respectively.}
\label{fig:den_con}
\label{fig:den_con}
\end{figure*}

Fig. \ref{fig:den_con} shows the results. The left- and right-hand-side are the results of GSPH and
the standard SPH, respectively. In the SPH simulation, the coefficients of the artificial viscosity
(they are called  $\displaystyle\alpha \text{ and }\beta$ in general)
are set to 1 and 2, respectively. Any other kind of treatment,
e.g., the artificial conduction \citep{Price2008a} or Balsala switch \citep{Balsara1995a} has not been
used.
The red solid lines are the analytic solution without hydrodynamics
given by Eqs. \eqref{eq:ana_up} - \eqref{eq:ana_low}. The solid circles and crosses
are the upper and lower layer material, respectively.
Apparently, there is no significant difference, but the numerical dispersion
of GSPH looks less serious, especially, at the border of the two layers
in the first and last snapshots.
It may be understandable, because the two layers are parallel,
so the curvature becomes 0. The numerical surface tension is
very similar to the physical one, so the higher curvature will make the stronger surface tension
\citep{Cha2010a}. One may see a clearer difference in the results, if any other curved geometry
(e.g. cylindrical fluid) has been employed.

\section{Viscous ring evolution}
\label{sec:viscousring}
The Couette flow is a good test problem for the viscous fluid simulations due to its simplicity
and clear analytic description. However, we need a more realistic test problem to check
our implementation of the physical viscosity.
\citet{Lynden-Bell1974a} derived the self-similar solution for the dynamical evolution of a viscous
gas ring under a central gravity, and it becomes a good test for a numerical code of the shear viscosity
\citep[e.g.][]{Flebbe1994a}. Only a brief description of the problem is presented here, because
there are many good sources that explain this problem
\citep[e.g.][]{Pringle1981a, Frank2002a}.

A ring of gas is located around the central gravitational object. The total mass and the
initial radius
of the ring are set to $m$ and $r_o$, respectively. The ring rotates around the central object,
and spreads due to the kinematic viscous, $\nu$. A dimensionless time,
\begin{align}
\tau = \frac{12\nu t}{r_o^2}
\end{align}
and radius,
\begin{align}
x = \frac{r}{r_o}
\end{align}
are useful to describe the evolution of the ring. The time evolution of the
surface density of the ring is given by
\begin{align}
\Sigma(\tau, x) = \left(\frac{m}{\pi r_o^2}\right)\frac{1}{\tau x^\frac{1}{4}}
\exp\left[-\frac{1+x^2}{\tau}\right]
I_\frac{1}{4}(z),
\end{align}
where $\displaystyle I_\frac{1}{4}$ and $\displaystyle z$ are the modified Bessel function and
$\displaystyle 2x/\tau$, respectively.
Furthermore, the radial velocity of the ring is given by
\begin{align}
\label{eq:vr_ring}
v_r (\tau,x)= \frac{6\nu}{\tau r_o}\left[x-\frac{I_{-\frac{3}{4}}(z)}{I_\frac{1}{4}(z)}\right].
\end{align}

The initial distribution of the ring material is set to the delta function,
\begin{align}
\Sigma(0,r) = \frac{m}{2\pi r_o} \delta(r-r_o)
\end{align}
in the analytic work of \citet{Lynden-Bell1974a}, but we choose the solution at
$\tau=1.6\times10^{-2}$ as the initial condition. Note that the radial velocity of our initial condition
should be defined by Eq. \eqref{eq:vr_ring} rather than $0$.
Initially, the mass and radius of the ring are set to 1, and the kinematic viscosity, $\nu$ is set to
$3\times 10^{-3}$ or $5\times 10^{-3}$. The two--dimensional GSPH code has been
used, and the total number of particles in the simulations is 10410.

\begin{figure*}
\centering
\includegraphics[width=0.4\textwidth]{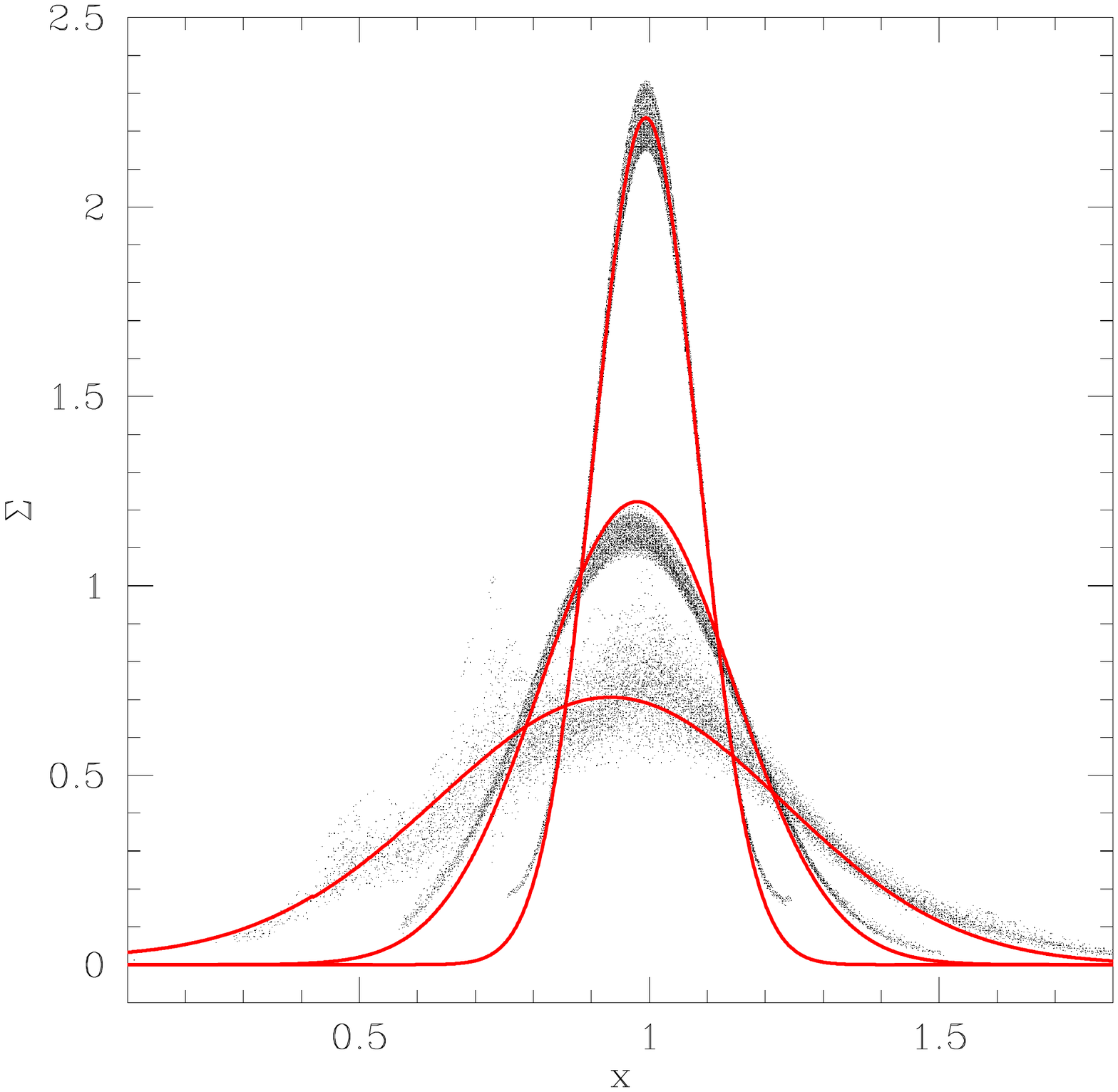}
\includegraphics[width=0.4\textwidth]{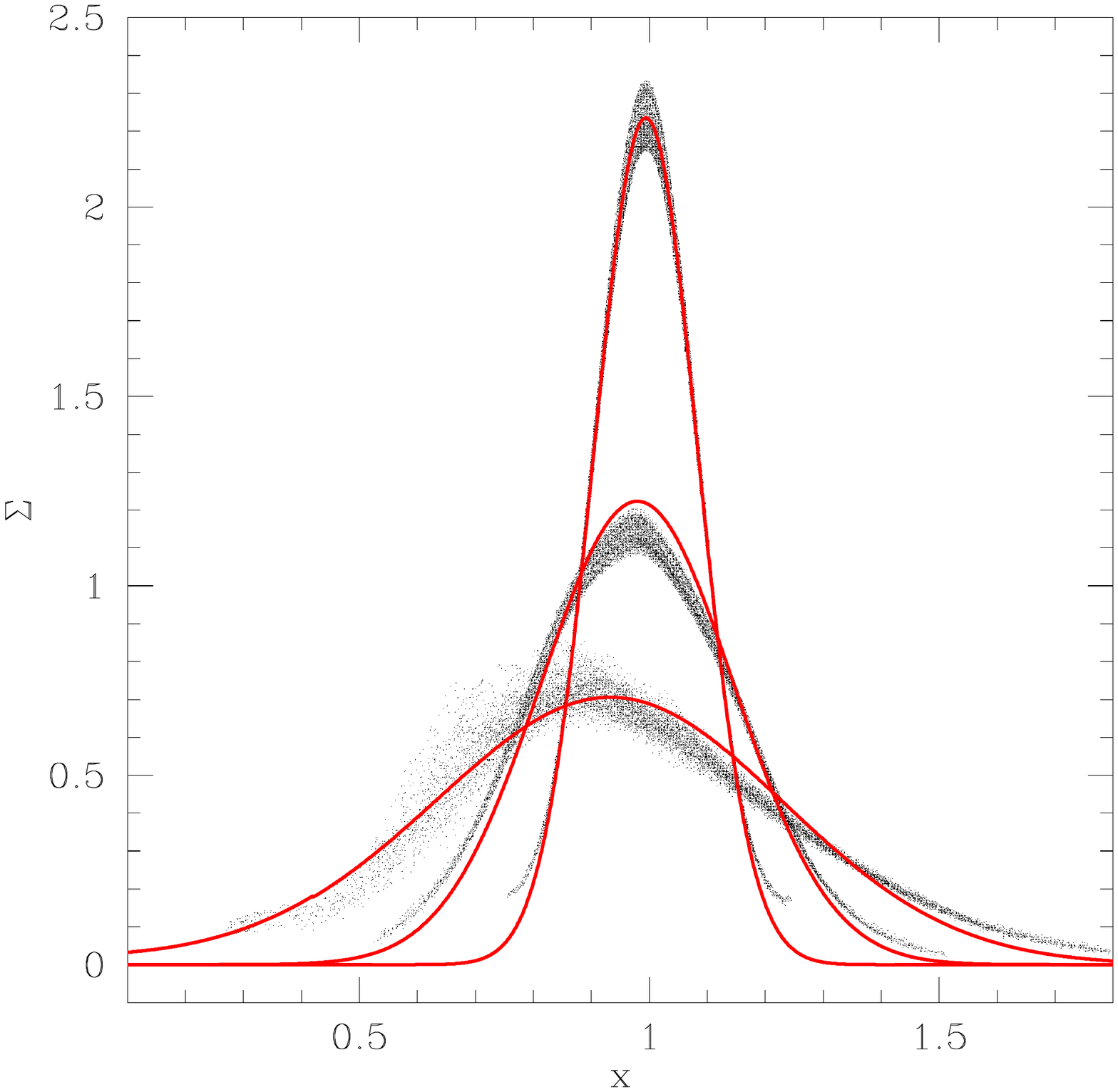}
\caption{The surface density of viscous ring evolution test.
The left and right panels are the result with  $\nu=3\times10^{-3}$ and $5\times10^{-3}$,
respectively, The evolution times are at $\tau=0.016,\,\,0.054\,\,\text{and}\,\,0.168$. The red solid lines
are the analytic solution, and the dots are the surface density of the ring. The overall agreement are
good, but the evolution is little bit faster than the analytic solution. Furthermore, the smaller kinematic viscosity
shows the more dispersive pattern.}
\label{fig:vis_ring}
\end{figure*}

Fig. \ref{fig:vis_ring} shows the results. The left and right panels are
$\nu=3\times 10^{-3}$ and $5\times 10^{-3}$, respectively. The red solid lines are
the analytic solution at $\tau=0.016,\,\,0.054\,\,\text{and}\,\,0.168$, and
the black dots are the surface density of the numerical calculations.
Note that the snapshot at $\tau=0.016$ is
the initial condition. The results shows an agreement to the analytic solution, but
a dispersive pattern happens in the late evolution stage of the small kinematic viscosity case
\citep{Speith1999a,Iwasaki2015a,Sugiura2016a}. Furthermore, the numerical results both are faster more or less than
the analytic solution \citep{Watkins1996a}.

\section{Conlusion and discussion}
\label{sec:conclusion}
Although there are many advantages of SPH as a multidimensional Lagrangian
code, the side effect of the artificial viscosity is a great obstacle to
simulate the accretion disks in CVs and around a protostar. The exact contribution
of the physical shear viscosity in the standard SPH
is hardly managed due to the interference of the
artificial viscosity. 

We have implemented several numerical routines to manage the viscous acceleration
and energy dissipation in our GSPH code. Contrary to the standard SPH, GSPH uses
the Riemann solver to describe the fluid motion and energy dissipation, therefore, it is free from
the side effect of the artificial viscosity. The double summation method has been used
in the treatment of the viscosity stress tensor and its derivatives. The second
derivative of the velocity can be described successfully, and the numerical noise
suppressed by this method.

The plane Couette flow has been chosen as a test problem. Our implementation shows
a good agreement with the analytic solution.  We have performed
not only the viscosity--only simulations, but also
the full hydrodynamics simulations of the viscous fluid. The hydrodynamics simulations
also show an expected results.

Simulations with a high $Re$ value have been performed as well.
We can see a dispersive pattern around
$Re \simeq 100$, and it is thought to be an occurrence of the turbulence. An experimental
and analytic expectation of $Re$ value for the turbulence occurrence
in the Couette flow is $325$ approximately,
so our simulation looks more sensitive against the occurrence of the turbulence.

In general, the kinematic viscosity is not a constant all over the calculation domain, and is a
function of the physical properties of a fluid element. The smoothed dynamic viscosity
has been introduced to describe the individual viscosity of particles. Simulations
with two layers which have different kinematic viscosities have been performed,
and the results show a good agreement
to the analytic solution, even in a high contrast of the kinetic viscosity.

The same two layers simulation has been formed with a density contrast as well.
The result of GSPH shows a less dispersive pattern at the boundary of the different density layers,
but the results of the standard SPH and GSPH does not show any meaningful difference.
It is because the numerical surface tension is not very critical due to the plane geometry.

The viscous gas ring simulation has been performed. The numerical solution can follow
the analytic solution well, but one can see a dispersive pattern at the later stage of the
small viscosity case. Furthermore, the evolution shows a faster evolution
than the analytic solution more or less.

We do not claim that GSPH is a complete inviscid code. The Riemann solver generates
a non--zero dissipation to describe shockwaves. However, it is hard to estimate
how much the dissipation is in the Godunov method \citep{Dunhill2013a,Puri2014a}. Perhaps,
the intrinsic numerical viscosity of GSPH should be added to the physical viscosity
somehow. However, we do not expect the mixing between the numerical and physical
viscosities to be a simple arithmetic sum.
We guess it may be the reason why the critical $Re$ value between the laminar and
turbulent flows in our simulations is different to the previous study. 
The exact contribution of the intrinsic viscosity of GSPH, and a better understanding
for the occurrence of the turbulence are left for a future study.

\section*{Acknowledgments}
We thank Anthony Whitworth and Andrew King for useful discussion on the selection of the test
problem. This material is based upon work supported by the National
Science Foundation under Grant No. AST-1305799 to Texas
A\&M University-Commerce and by NASA under grant 11-KEPLER11-0038.

\bibliographystyle{mn2e}
\bibliography{mnemonic,shcha}

\begin{thebibliography}{}

\bibitem[\protect\citeauthoryear{{Agertz}, {Moore}, {Stadel}, {Potter},
  {Miniati}, {Read}, {Mayer}, {Gawryszczak}, {Kravtsov}, {Nordlund}, {Pearce},
  {Quilis}, {Rudd}, {Springel}, {Stone}, {Tasker}, {Teyssier}, {Wadsley} \&
  {Walder}}{{Agertz} et~al.}{2007}]{Agertz2007a}
{Agertz} O.,  {Moore} B.,  {Stadel} J.,  {Potter} D.,  {Miniati} F.,  {Read}
  J.,  {Mayer} L.,  {Gawryszczak} A.,  {Kravtsov} A.,  {Nordlund} {\AA}.,
  {Pearce} F.,  {Quilis} V.,  {Rudd} D.,  {Springel} V.,  {Stone} J.,  {Tasker}
  E.,  {Teyssier} R.,  {Wadsley} J.,    {Walder} R.,  2007, MNRAS, 380, 963

\bibitem[\protect\citeauthoryear{{Artymowicz} \& {Lubow}}{{Artymowicz} \&
  {Lubow}}{1994}]{Artymowicz1994a}
{Artymowicz} P.,  {Lubow} S.~H.,  1994, ApJ, 421, 651

\bibitem[\protect\citeauthoryear{{Balsara}}{{Balsara}}{1995}]{Balsara1995a}
{Balsara} D.~S.,  1995, J. Comp. Phys., 121, 357

\bibitem[\protect\citeauthoryear{{Barnes} \& {Hut}}{{Barnes} \&
  {Hut}}{1986}]{Barnes1986a}
{Barnes} J.,  {Hut} P.,  1986, Nature, 324, 446

\bibitem[\protect\citeauthoryear{{Brookshaw}}{{Brookshaw}}{1985}]{Brookshaw1985a}
{Brookshaw} L.,  1985, PASA, 6, 207

\bibitem[\protect\citeauthoryear{{Carslaw} \& {Jaeger}}{{Carslaw} \&
  {Jaeger}}{1959}]{Carslaw1959a}
{Carslaw} H.~S.,  {Jaeger} J.~C.,  1959, {Conduction of heat in solids}.
Oxford University Press

\bibitem[\protect\citeauthoryear{{Cha}, {Inutsuka} \& {Nayakshin}}{{Cha}
  et~al.}{2010}]{Cha2010a}
{Cha} S.-H.,  {Inutsuka} S.-I.,    {Nayakshin} S.,  2010, MNRAS, 403, 1165

\bibitem[\protect\citeauthoryear{{Cha} \& {Whitworth}}{{Cha} \&
  {Whitworth}}{2003}]{Cha2003a}
{Cha} S.-H.,  {Whitworth} A.~P.,  2003, MNRAS, 340, 73

\bibitem[\protect\citeauthoryear{{Chaniotis}, {Poulikakos} \&
  {Koumoutsakos}}{{Chaniotis} et~al.}{2002}]{Chaniotis2002a}
{Chaniotis} A.~K.,  {Poulikakos} D.,    {Koumoutsakos} P.,  2002, J. Comp.
  Phys., 182, 67

\bibitem[\protect\citeauthoryear{{Cleary} \& {Monaghan}}{{Cleary} \&
  {Monaghan}}{1999}]{Cleary1999a}
{Cleary} P.~W.,  {Monaghan} J.~J.,  1999, J. Comp. Phys., 148, 227

\bibitem[\protect\citeauthoryear{{Couette}}{{Couette}}{1890}]{Couette1890a}
{Couette} M.,  1890, Ann. de Chimie et de Physique, 21, 433

\bibitem[\protect\citeauthoryear{{Dunhill}, {Alexander} \&
  {Armitage}}{{Dunhill} et~al.}{2013}]{Dunhill2013a}
{Dunhill} A.~C.,  {Alexander} R.~D.,    {Armitage} P.~J.,  2013, MNRAS, 428,
  3072

\bibitem[\protect\citeauthoryear{{Fatehi}, {Fayazbakhsh} \& {Manzari}}{{Fatehi}
  et~al.}{2009}]{Fatehi2009a}
{Fatehi} R.,  {Fayazbakhsh} M.,    {Manzari} M.~T.,  2009, Int. Jour. Nat. and
  App. Sci, 3, 50

\bibitem[\protect\citeauthoryear{{Flebbe}, {Muenzel}, {Herold}, {Riffert} \&
  {Ruder}}{{Flebbe} et~al.}{1994}]{Flebbe1994a}
{Flebbe} O.,  {Muenzel} S.,  {Herold} H.,  {Riffert} H.,    {Ruder} H.,  1994,
  ApJ, 431, 754

\bibitem[\protect\citeauthoryear{{Frank}, {King} \& {Raine}}{{Frank}
  et~al.}{2002}]{Frank2002a}
{Frank} J.,  {King} A.,    {Raine} D.~J.,  2002, {Accretion Power in
  Astrophysics: Third Edition}.
Cambridge University Press

\bibitem[\protect\citeauthoryear{{Gingold} \& {Monaghan}}{{Gingold} \&
  {Monaghan}}{1977}]{Gingold1977a}
{Gingold} R.~A.,  {Monaghan} J.~J.,  1977, MNRAS, 181, 375

\bibitem[\protect\citeauthoryear{{Godunov}}{{Godunov}}{1959}]{Godunov1959a}
{Godunov} S.~K.,  1959, Math. Sbornik, 47, 271

\bibitem[\protect\citeauthoryear{{Inutsuka}}{{Inutsuka}}{2002}]{Inutsuka2002a}
{Inutsuka} S.-I.,  2002, J. Comp. Phys., 179, 238

\bibitem[\protect\citeauthoryear{{Iwasaki}}{{Iwasaki}}{2015}]{Iwasaki2015a}
{Iwasaki} K.,  2015, J. Comp. Phys., 302, 359

\bibitem[\protect\citeauthoryear{{Iwasaki} \& {Inutsuka}}{{Iwasaki} \&
  {Inutsuka}}{2011}]{Iwasaki2011a}
{Iwasaki} K.,  {Inutsuka} S.-I.,  2011, MNRAS, 418, 1668

\bibitem[\protect\citeauthoryear{{Kitoh}, {Nakabyashi} \& {Nishimura}}{{Kitoh}
  et~al.}{2005}]{Kitoh2005a}
{Kitoh} O.,  {Nakabyashi} K.,    {Nishimura} F.,  2005, Jour. Fluid Mech., 539,
  199

\bibitem[\protect\citeauthoryear{{Lanzafame}}{{Lanzafame}}{2003}]{Lanzafame2003a}
{Lanzafame} G.,  2003, A\&A, 403, 593

\bibitem[\protect\citeauthoryear{{Lubow}}{{Lubow}}{1991}]{Lubow1991a}
{Lubow} S.~H.,  1991, ApJ, 381, 268

\bibitem[\protect\citeauthoryear{{Lucy}}{{Lucy}}{1977}]{Lucy1977a}
{Lucy} L.~B.,  1977, AJ, 82, 1013

\bibitem[\protect\citeauthoryear{{Lynden-Bell} \& {Pringle}}{{Lynden-Bell} \&
  {Pringle}}{1974}]{Lynden-Bell1974a}
{Lynden-Bell} D.,  {Pringle} J.~E.,  1974, MNRAS, 168, 603

\bibitem[\protect\citeauthoryear{{Monaghan}}{{Monaghan}}{1992}]{Monaghan1992a}
{Monaghan} J.~J.,  1992, ARA\&A, 30, 543

\bibitem[\protect\citeauthoryear{{Monaghan}}{{Monaghan}}{2005}]{Monaghan2005b}
{Monaghan} J.~J.,  2005, Rep. Prog. Phys., 68, 1703

\bibitem[\protect\citeauthoryear{{Price}}{{Price}}{2008}]{Price2008a}
{Price} D.~J.,  2008, J. Comp. Phys., 227, 10040

\bibitem[\protect\citeauthoryear{{Pringle}}{{Pringle}}{1981}]{Pringle1981a}
{Pringle} J.~E.,  1981, ARA\&A, 19, 137

\bibitem[\protect\citeauthoryear{{Puri} \& {Ramachandran}}{{Puri} \&
  {Ramachandran}}{2014}]{Puri2014a}
{Puri} K.,  {Ramachandran} P.,  2014, J. Comp. Phys., 270, 432

\bibitem[\protect\citeauthoryear{{Schneider}, {de Lillo}, {Buehrle},
  {Eckhardt}, {D{\"o}rnemann}, {D{\"o}rnemann} \& {Freisleben}}{{Schneider}
  et~al.}{2010}]{Schneider2010a}
{Schneider} T.~M.,  {de Lillo} F.,  {Buehrle} J.,  {Eckhardt} B.,
  {D{\"o}rnemann} T.,  {D{\"o}rnemann} K.,    {Freisleben} B.,  2010, Phys.
  Rev. E, 81, 015301

\bibitem[\protect\citeauthoryear{{Shakura} \& {Sunyaev}}{{Shakura} \&
  {Sunyaev}}{1973}]{Shakura1973a}
{Shakura} N.~I.,  {Sunyaev} R.~A.,  1973, A\&A, 24, 337

\bibitem[\protect\citeauthoryear{{Speith} \& {Riffert}}{{Speith} \&
  {Riffert}}{1999}]{Speith1999a}
{Speith} R.,  {Riffert} H.,  1999, J. Comp. Appl. Math., 109, 231

\bibitem[\protect\citeauthoryear{{Sugiura} \& {Inutsuka}}{{Sugiura} \&
  {Inutsuka}}{2016}]{Sugiura2016a}
{Sugiura} K.,  {Inutsuka} S.-i.,  2016, J. Comp. Phys., 308, 171

\bibitem[\protect\citeauthoryear{{Takeda}, {Miyama} \& {Sekiya}}{{Takeda}
  et~al.}{1994}]{Takeda1994a}
{Takeda} H.,  {Miyama} S.~M.,    {Sekiya} M.,  1994, Prog. Theor. Phys, 92, 939

\bibitem[\protect\citeauthoryear{{Tsukamoto}, {Iwasaki} \&
  {Inutsuka}}{{Tsukamoto} et~al.}{2013}]{Tsukamoto2013a}
{Tsukamoto} Y.,  {Iwasaki} K.,    {Inutsuka} S.-i.,  2013, MNRAS, 434, 2593

\bibitem[\protect\citeauthoryear{{Tsukamoto}, {Iwasaki}, {Okuzumi}, {Machida}
  \& {Inutsuka}}{{Tsukamoto} et~al.}{2015}]{Tsukamoto2015a}
{Tsukamoto} Y.,  {Iwasaki} K.,  {Okuzumi} S.,  {Machida} M.~N.,    {Inutsuka}
  S.,  2015, MNRAS, 452, 278

\bibitem[\protect\citeauthoryear{{van Leer}}{{van Leer}}{1997}]{vanLeer1997a}
{van Leer} B.,  1997, J. Comp. Phys., 135, 229

\bibitem[\protect\citeauthoryear{{Viau}, {Bastien} \& {Cha}}{{Viau}
  et~al.}{2006}]{Viau2006a}
{Viau} S.,  {Bastien} P.,    {Cha} S.-H.,  2006, ApJ, 639, 559

\bibitem[\protect\citeauthoryear{{Watkins}, {Bhattal}, {Francis}, {Turner} \&
  {Whitworth}}{{Watkins} et~al.}{1996}]{Watkins1996a}
{Watkins} S.~J.,  {Bhattal} A.~S.,  {Francis} N.,  {Turner} J.~A.,
  {Whitworth} A.~P.,  1996, A\&AS, 119, 177

\bibitem[\protect\citeauthoryear{{Wood} \& {Burke}}{{Wood} \&
  {Burke}}{2007}]{Wood2007a}
{Wood} M.~A.,  {Burke} C.~J.,  2007, ApJ, 661, 1042

\bibitem[\protect\citeauthoryear{{Wood}, {Thomas} \& {Simpson}}{{Wood}
  et~al.}{2009}]{Wood2009a}
{Wood} M.~A.,  {Thomas} D.~M.,    {Simpson} J.~C.,  2009, MNRAS, 398, 2110

\end{thebibliography}



\bsp

\label{lastpage}

\end{document}